\documentclass[usenatbib]{mnras}
\usepackage{graphicx}
\usepackage{natbib}
\usepackage{latexsym}
\usepackage{amsmath}
\usepackage{setspace}
\usepackage{booktabs}
\usepackage{array, longtable}
\usepackage[bf, singlelinecheck=off]{caption}
\usepackage{subfig}
\usepackage[T1]{fontenc}
\usepackage{aecompl}

\newcommand{\etal}{et~al.} 
\newcommand{\ionhy}{H{\sc ii} }

\newcommand{\kms}{$\mbox{km~s}^{-1}$}

\newcommand{\specsfig}[1]        
{
   \begin{center}
     \begin{minipage}[t]{0.45\textwidth}
         \psfig{file=#1.eps,height=0.9\textwidth,angle=270}
     \end{minipage}
     \end{center}
 }

\newcommand{\specdfig}[2]        
{
   \begin{center}
     \begin{minipage}[t]{0.45\textwidth}
         \psfig{file=#1.eps,height=0.9\textwidth,angle=270}
     \end{minipage}
     \hfill
     \begin{minipage}[t]{0.45\textwidth}
         \psfig{file=#2.eps,height=0.9\textwidth,angle=270}
     \end{minipage}
   \end{center}
}

\def \la{\mathrel{\mathchoice   {\vcenter{\offinterlineskip\halign{\hfil
					$\displaystyle##$\hfil\cr<\cr\sim\cr}}}
		{\vcenter{\offinterlineskip\halign{\hfil$\textstyle##$\hfil\cr
					<\cr\sim\cr}}}
		{\vcenter{\offinterlineskip\halign{\hfil$\scriptstyle##$\hfil\cr
					<\cr\sim\cr}}}
		{\vcenter{\offinterlineskip\halign{\hfil$\scriptscriptstyle##$\hfil\cr
					<\cr\sim\cr}}}}}
\def \ga{\mathrel{\mathchoice   {\vcenter{\offinterlineskip\halign{\hfil
					$\displaystyle##$\hfil\cr>\cr\sim\cr}}}
		{\vcenter{\offinterlineskip\halign{\hfil$\textstyle##$\hfil\cr
					>\cr\sim\cr}}}
		{\vcenter{\offinterlineskip\halign{\hfil$\scriptstyle##$\hfil\cr
					>\cr\sim\cr}}}
		{\vcenter{\offinterlineskip\halign{\hfil$\scriptscriptstyle##$\hfil\cr
					>\cr\sim\cr}}}}}

\begin{document}
\title{Investigations of the Class I methanol masers in NGC~4945}
\author[McCarthy \etal]{
	T.\ P. McCarthy,$^{1,2}$\thanks{Email: tiegem@utas.edu.au}
	S.\ P. Ellingsen,$^{1}$
    S. L. Breen,$^{3}$
    C. Henkel$^{4,5}$ 
	M.\ A. Voronkov$^{2}$ and
	\newauthor X. Chen$^{6,7}$ \\
  \\
  $^1$ School of Natural Sciences, University of Tasmania, Private Bag 37, Hobart, Tasmania 7001, Australia\\
  $^2$ Australia Telescope National Facility, CSIRO, PO Box 76, Epping, NSW 1710, Australia \\
  $^3$ Sydney Institute for Astronomy (SIfA), School of Physics, University of Sydney, NSW 2006, Australia\\
  $^4$ Max-Planck-Institut f\"ur Radioastronomie, Auf dem H\"ugel 69, 53121 Bonn, Germany \\
  $^5$ Astronomy Department, Faculty of Science, King Abdulaziz University, P.O. Box 80203, Jeddah 21589, Saudi Arabia\\
  $^6$ Center for Astrophysics, GuangZhou University, Guangzhou 510006, China\\
  $^7$ Shanghai Astronomical Observatory, Chinese Academy of Sciences, Shanghai 200030, China \\}

 \maketitle

\begin{abstract}

We have used the Australia Telescope Compact Array (ATCA) to conduct further observations of the 36.2-GHz ($4_{-1}\rightarrow3_0$E) methanol transition towards the nearby active galaxy NGC~4945. These observations have led to a more accurate determination of the offset between the maser emission and the nucleus of NGC~4945 with a typical synthesised beam of $6^{\prime\prime} \times 4^{\prime\prime}$ ($108\times72$ pc). This corresponds to a factor of 4 improvement with respect to the major-axis of  the beam. Other transitions of methanol and lines of other molecular species were obtained alongside the 36.2-GHz methanol emission, with strong detections of HC$_3$N~(J~=~$4 \rightarrow 3$) and CS~(J~=~$1 \rightarrow0$) presented here. We do not detect thermal methanol (5$\sigma$ upper limit of 5~mJy in a 6~\kms\ channel) from the 48.4-GHz ($1_{0}\rightarrow0_0$A$^+$) ground-state transition, nor emission from the 44.1-GHz ($7_{0} \rightarrow 6_1 $A$^+$) class~I maser transition (5$\sigma$ upper limit of 6 mJy in a 3~\kms\ channel). We also present a comparison of the class~I maser emission observed towards NGC~4945 with that from NGC~253 and towards the Galactic giant molecular cloud G~1.6-0.025.

\end{abstract}


\section{Introduction}

\begin{table*}
	\begin{center}
		\caption{Details of the observed transitions in NGC~4945. All rest frequencies were taken from the online database: \textit{NIST Recommended Rest Frequencies for Observed Interstellar Molecular Microwave Transitions
			by Frank J. Lovas}\protect\footnote[1]{Test 1, 2 3}.}
		\begin{tabular}{llcllcl} \hline
			\multicolumn{1}{c}{\bf Species} & \multicolumn{1}{c}{\bf Transition} &  \multicolumn{1}{c}{\bf Rest} & \multicolumn{1}{c}{\bf Array}  & \multicolumn{1}{c}{\bf Epoch} & \multicolumn{1}{c}{\bf Integration}  &\multicolumn{1}{c}{\bf Detection} \\
			& & \multicolumn{1}{c}{\bf Frequency} & \multicolumn{1}{c}{\bf Configuration}& & \multicolumn{1}{c}{\bf Time} & \\  
			& & \multicolumn{1}{c}{(GHz)} & & & \multicolumn{1}{c}{(min)} & \\
			\hline
			Methanol & $4_{-1} \rightarrow 3_0 $E & 36.169265 &  EW352, H168, H214 & 2015 Aug, 2017 Jun, 2017 Oct & 265.2 & detection  \\    
			& $7_{-2} \rightarrow 8_{-1} $E & 37.703696 & EW352, H168, H214  & 2015 Aug, 2017 Jun, 2017 Oct & 265.2 & non-detection  \\    
			& $7_{0} \rightarrow 6_1 $A$^+$ & 44.069476 & H214  & 2017 Oct & 178.2 & non-detection  \\    
			& $1_{0} \rightarrow 0_0 $A$^+$ & 48.372456 & H214  & 2017 Oct & 138.2 & non-detection   \\    
			HC$_3$N & $J=4 \rightarrow 3~F = 4 \rightarrow 3$ & 36.392332 & H168, H214 & 2017 Jun, 2017 Oct & 197.4 & detection  \\
			HC$_5$N & $J=12 \rightarrow 11$ & 31.951777 & H168, H214 & 2017 Jun, 2017 Oct & 197.4 & non-detection  \\
			NH$_3$ & $12_{12} \rightarrow 12_{12}$ & 31.424943 & H168, H214 & 2017 Jun, 2017 Oct & 197.4 & non-detection  \\
			& $13_{13} \rightarrow 13_{13}$ & 33.156849 & H168, H214 & 2017 Jun, 2017 Oct & 197.4 & non-detection  \\
			CH$_3$CN & $2_{0} \rightarrow 1_{0},~F=3\rightarrow2$ & 36.795568 & H168, H214 & 2017 Jun, 2017 Oct & 98.4 & non-detection  \\
			SiO & $J=1 \rightarrow 0,~\text{v}=0$ & 43.423853 & H214 & 2017 Oct & 178.2 & detection  \\
			CS & $J=1\rightarrow0$ & 48.990955 & H214 & 2017 Oct & 138.2 & detection  \\  \hline
		\end{tabular} \label{tab:observations}		
	\end{center}
\end{table*}

Methanol maser emission is divided into two classes based on pumping mechanism. Methanol masers pumped via collisional processes are defined as class~I, while those that are radiatively pumped are considered class~II \citep{Batrla+87, Menten91a}. Both classes of methanol masers are commonly observed throughout the Milky Way, with over 1200 unique sources discovered \citep[e.g][]{Ellingsen+05,Caswell+10,Caswell+11, Voronkov+14, Breen+15, Green+10, Green+12a, Green+17}. In contrast with this, outside of our Galaxy we have relatively few detections of methanol maser emission. Extragalactic class~II masers have been detected in the Large Magellanic Cloud and M31 \citep{Green+08,Ellingsen+10,Sjouwerman+10}. These extragalactic class~II masers appear to be extremely luminous examples of their Galactic counterparts. Conversely, the extragalactic class~I methanol masers are not yet a well understood phenomenon, with observed emission unable to simply be considered large-scale emission from Galactic-style class~I masers. Currently there are 6 reported examples of class~I maser emission towards extragalactic sources, 36.2-GHz emission in NGC~253, Arp~220, IC~342, NGC~6946 and NGC~4945 \citep{Ellingsen+14, Chen+15, McCarthy+17, Gorski+18} and 84.5-GHz emission in NGC~1068 \citep{Wang+14}. Of these 6 sources, only NGC~253 has been detected in multiple epochs with multiple telescopes \citep{Ellingsen+14,Ellingsen+17b,Chen+18,Gorski+18}. 

Class~I methanol emission is a powerful tool for understanding star-formation within our Galaxy. More than 600 unique sources of class~I maser emission are observed within the Milky Way \citep[e.g.,][]{Slysh+94,Valtts+00,Ellingsen+05,Chen+11,Gan+13,Jordan+17}. Galactic class~I masers are generally associated with shocked gas driven by the expansion of \ionhy regions or molecular outflows \citep{Kurtz+04, Cyganowski+09, Cyganowski+12, Voronkov+10a, Voronkov+14}. However, it is not yet known if, or how, these highly luminous extragalactic class~I masers relate to the star-formation of their host galaxies. In Galactic star formation regions the two most commonly observed class~I methanol maser transitions are the 36.2- and the 44.1-GHz, with the latter generally being the stronger of the two \citep[e.g. ][]{Voronkov+14}.  \citep{Ellingsen+17b} detected weak 44.1-GHz methanol maser emission associated with two of the 36.2-GHz sites in NGC\,253.  They suggest that the low intensity of the 44.1-GHz transition compared to the 36.2-GHz in NGC\,253 is strong evidence that extragalactic class~I methanol maser emission cannot be explained as being due a large number of Galactic-like star formation regions within a small volume, but is rather a new and different extragalactic masing phenomenon. It appears that the extragalactic variants may evolve from large-scale molecular inflow inside their host galaxies \citep{Ellingsen+17b}. However, this has so far only been verified in the case of NGC~253. Developing an understanding of the pumping environments responsible for this phenomenon is one of the most important factors in determining its usefulness as a probe of galactic properties.

NGC~4945 is a nearby \citep[assumed distance of 3.7$\pm0.3$\,Mpc;][]{Tully+13} spiral galaxy, with a hybrid AGN and starburst nucleus. The starburst is the primary source of energy for exciting photo-ionized gas, due to heavy obscuration of the AGN by dust \citep{Marconi+00, Spoon+00, Spoon+03, Perez-Beauputis+11}. The star-formation rate in NGC~4945 is more than three times that of the Milky Way \citep[$4.35\pm0.25$ M$_{\sun}$ yr$^{−1}$ for the nuclear region of NGC~4945 only, compared to $1.65\pm0.19$ M$_{\sun}$ yr$^{−1}$ for the entire Milky Way;][]{Bendo+16,Licquia+15} and approximately 20 percent higher than that of the similar (in terms of galactic properties and maser luminosity) extragalactic class~I maser host galaxy, NGC~253 \citep{Strickland+04}. NGC~4945 is also host to various transitions of water megamasers, predominantly located in a circumnuclear accretion disk \citep{Greenhill+97, Hagiwara+16,Humphreys+16, Pesce+16}.

We have undertaken new observations of the 36.2-GHz methanol maser transition in NGC\,4945 to better determine its location with respect to the host galaxy and other molecular gas.  In addition to the 36.2-GHz methanol transition we have also observed the 44.1-GHz class~I methanol maser transition to determine if the NGC\,4945 shows a similar pattern to NGC\,253 with this transition being relatively much weaker than is observed towards Galactic class~I methanol masers associated with high-mass star formation regions.  We were able to include observations of a number of thermal molecular transitions simultaneously with the maser observations and we present the results of those observations and compare them with the recent, sensitive high-resolution molecular line ALMA observations at 3-mm made by \citet{Henkel+18}.  We currently have a sample of only six known extragalactic class~I methanol maser sources and by obtaining a range of complementary spectral line and other data and comparing the results for NGC\,4945 with other sources we hope to improve understanding of this new phenomenon and its relationship to the properties of the host galaxy.

\footnotetext[1]{https://physics.nist.gov/cgi-bin/micro/table5/start.pl}


\section{Observations} \label{sec:observations}

The Australia Telescope Compact Array (ATCA) was utilised for observations of NGC~4945 on 2017 June 29 and 2017 October 22 (project code C3167). The observations both used hybrid array configurations, H214 for the June session (minimum and maximum baselines of 82 and 247m respectively) and H168 for the October session (minimum and maximum baselines of 61 and 192m respectively), and we excluded antenna 6 from our analysis. The Compact Array Broadband Backend \citep[CABB ;][]{Wilson+11} was configured in CFB 64M-32k mode for these sessions. This mode consists of two 2 GHz IF bands (consisting of $32 \times 64$~MHz channels) and up to 16 of these 64~MHz channels can be configured as zoom bands consisting of $2048 \times 31.2$~kHz channels. The October observing session consisted of two separate frequency setups, splitting the transitions below and above 40 GHz. Multiple zoom bands were `stitched' together in order to obtain the appropriate velocity coverage for each transition. Table \ref{tab:observations} describes all the molecular transitions, along with which epochs they were observed in. The two class I maser lines at 36.2- ($4_{-1} \rightarrow 3_0 E$) and 44.1-GHz ($7_{0} \rightarrow 6_1 A^+$) were the primary science targets, for which we adopted rest frequencies of 36.169265 and 44.069410 GHz respectively. The 31.2~kHz spectral resolution corresponds to 0.259~\kms and 0.213~\kms\ at 36.2 and 44.1 GHz respectively.

The following refers to the observing strategy for the 2017 June and October epochs. Details regarding the previously reported 2015 August observations (EW352 array configuration) can be found in \citet{McCarthy+17}. PKS~B1934-648 was used for flux density calibration for both epochs and the bandpass was calibrated with respect to PKS~B0537-441 and PKS~B1921-293. PMN~J1326-5256 was utilised as the phase calibrator, with 2 minutes on the calibrator interleaved with 10 minutes on-source. The data were corrected for atmospheric opacity and the absolute flux density calibration is estimated to be accurate to better than 30\%. Poor weather during the latter half of the 2017 October observations negatively affected the transitions above 48~GHz to a more significant degree. This, combined with representing only single epoch data, causes a significantly higher RMS noise in these higher frequency transitions (see Table \ref{tab:cubes}).

{\sc miriad} was used for data reduction, following standard techniques for the reduction of ATCA 7-mm spectral line observations. Phase and amplitude self-calibration was performed on the data using the continuum emission from the core of NGC~4945. The continuum emission was subtracted from the self-calibrated uv-data with the uvlin task, which estimates the intensity on each baseline from the line-free spectral channels. This enables us to isolate any spectral line emission from continuum emission.  Data from all relevant epochs were combined prior to imaging (for on-source times see Table~\ref{tab:observations}). Molecular line emission in NGC\,4945 is observed between approximately 300 and 800\,km\,s$^{-1}$ in the LSR velocity reference frame \citep{Ott+01}. The velocity range of our imaging was dependent on the number of stitched zoom bands utilised for each observed transition. However, all observed transitions were covered at least in the 350 to 780~\kms\ range. The spectral line data for each transition was resampled and imaged with a variety of channel widths. Table \ref{tab:cubes} shows the velocity range, channel width and RMS per channel for a typical spectral line cube of each transition. Positions were determined using the {\sc miriad} task imfit, which reports the peak value and location of a two-dimensional Gaussian fit for the emission in a given velocity plane within the spectral line cube. This task was also utilised for reporting peak flux-density values for our emission, which may result in minor differences between the apparent flux-density seen in extracted spectra and those tabulated. Combining data from all three array configurations results in a typical synthesised beam size of $6^{\prime\prime} \times 4^{\prime\prime}$.

\section{Results} \label{sec:results}

\begin{table}
	\begin{center}
		\caption{Details of the final spectral line cubes for each molecular line observed. Image cubes were created out of all available data. Methanol and continuum imaging included previously reported August 2015 observations, whereas, other transitions use both 2017 epochs where applicable. All details given refer to self-calibrated image cubes. RMS noise value is per channel, of width given in the relevant column. All velocities are with respect to the barycentric reference frame.}
		\begin{tabular}{llccc} \hline
			\multicolumn{1}{c}{\bf Species} & \multicolumn{1}{c}{\bf Rest} & \multicolumn{1}{c}{\bf Velocity}  & \multicolumn{1}{c}{\bf Channel} & \multicolumn{1}{c}{\bf RMS} \\
			\multicolumn{1}{c}{} & \multicolumn{1}{c}{\bf Frequency} & \multicolumn{1}{c}{\bf Range}  & \multicolumn{1}{c}{\bf Width} & \multicolumn{1}{c}{} \\
			& \multicolumn{1}{c}{(GHz)} & \multicolumn{1}{c}{(\kms)} & \multicolumn{1}{c}{(\kms)} & \multicolumn{1}{c}{(mJy)} \\  \hline
			Methanol & 36.169265  & 80 -- 840 & 1.0 & 2.0  \\    
			& 37.703696  & 200 -- 860  & 3.0  & 0.9  \\    
			& 44.069476  & 180 -- 810 & 3.0 & 1.2  \\    
			& 48.372456  & 150 -- 810  & 6.0 & 1.7   \\    
			HC$_3$N & 36.392332  & 200 -- 950 & 3.0 & 1.1  \\
			HC$_5$N & 31.951777 &  80 -- 980  & 3.0 & 0.7  \\
			NH$_3$ & 31.424943  & 250 -- 850 & 3.0 & 0.7  \\
			& 33.156849 & 270 -- 870 & 3.0 & 0.9  \\
			CH$_3$CN & 36.795568  & 350 -- 800 & 3.0 & 0.8  \\
			SiO & 43.423853  & 180 -- 780 & 3.0 & 1.3  \\
			CS & 48.990955  & 270 -- 930 & 3.0 & 3.4  \\  \hline
		\end{tabular} \label{tab:cubes}		
	\end{center}
\end{table}

The previously reported 36.2-GHz maser emission from NGC~4945 was clearly detected in both the June and October 2017 epochs, along with a strong 7-mm continuum source. In addition to methanol, 36.4-GHz HC$_3$N ($J = 4 \rightarrow 3$) line emission was detected in both epochs and 49-GHz CS $J = 1\rightarrow0$ and 43.4-GHz SiO $J = 1\rightarrow0,~\nu=0$ emission was also detected in the October session (where frequencies greater than 40 GHz were covered). Below we describe, in more detail, each of these detected transitions (listed in Table \ref{tab:observations}). All velocities referenced and reported in this work are relative to the barycentric coordinate system ($V_{\text{barycentric}}-V_{\text{LSR}}= 4.6~\text{\kms}$). Properties of the emission from detected transitions are tabulated in Table~\ref{tab:emission}.

\subsection{Continuum}

Using the imfit {\sc miriad} task on our combined array data gives a beam deconvolved angular size of $4.8\pm0.04 \times 1.8\pm0.04$ arcseconds, with a position angle of $41\fdg4\pm0\fdg3$ for the major axis. This is of similar size, and identical position angle, compared to the 3-mm continuum source dimensions reported by \citet{Henkel+18}.

The bulk of the nuclear continuum emission observed at millimetre wavelengths is the result of free-free emission from star formation regions \citep{Bendo+16}. We detect a 7-mm continuum source towards the nuclear region of NGC~4945, with peak and integrated flux-densities of 299~mJy and 436~mJy~\kms, respectively (see Table \ref{tab:emission} for position). This continuum emission was extracted by combining the line-free continuum data from all three epochs (and array configurations). The zoom band for the 36.2-GHz methanol line was used, in order to remain consistent with the previous value reported in \citet{McCarthy+17}. This zoom band covers the frequency range 35.99 -- 36.18 GHz. Our integrated flux density is in good agreement with the modelled values for 7-mm from \citet{Bendo+16}.  The SED model of \citeauthor{Bendo+16} (see their figure 5) predicts 75\% of the 7-mm emission is from free-free emission, so the good agreement we obtain suggests that at this frequency, the majority of the continuum emission is due to star formation.

The dynamical centre, as defined by the position of the circumnuclear H$_2$O megamaser accretion disk, is located at $\alpha_{2000} = 13^{\text{h}}~05^{\text{m}} ~ 27^{\text{s}}.48$ and $\delta_{2000} = -49^\circ ~ 28^\prime ~ 05\farcs6$ \citep{Greenhill+97, Henkel+18}. Comparing the peak of our 7-mm continuum source, we see an offset of $-0.13$ arcseconds in right ascension and $0.8$ arcseconds in declination (projected linear offset of 15~pc), indicating our continuum source is offset to the north-west from the dynamical centre. However, this offset is of the same order as the nominal astrometric accuracy of our observations, 0.4 arcseconds. The 3-mm continuum source detected by \citet{Henkel+18} with a beam size of $\sim2^{\prime\prime}$ shows a similar angular offset, though to the north-east, instead of north-west. We also identify an asymmetry between the angular offsets of the two CS $J = 1 \rightarrow 0$ peaks to the SW and NE when compared to the 7-mm continuum peak ($4\farcs4$ and $5\farcs7$ for the SW and NE components respectively). In addition to the main continuum source, we observe a minor emission feature (at a $6\sigma$ level) roughly aligned with the position of the maser emission to the south-east (see Figure \ref{fig:plot&spec}).

\subsection{Methanol spectral lines}\label{RES:Methanol}

The 36.2-GHz (class~I) methanol maser emission (see Table \ref{tab:emission}) is observed from the same position as reported in \citet{McCarthy+17} in both follow up epochs, offset south-east from the galactic nucleus. The hybrid array configuration used for both 2017 observations allows for more accurate imaging of the methanol masing region, which wasn't possible with previous observations. All emission is located at this offset region, which corresponds to the location of some weak HCN emission (see Figure \ref{fig:plot&spec}), and consists of multiple components with narrow line widths. The 55 mJy primary component has a velocity of 673~km\,s$^{-1}$, red-shifted with respect to the systemic velocity of NGC~4945 \citep[585~km\,s$^{-1}$;][]{Chou+07}, with two significantly lower flux density ($\sim$15 mJy) components at 643 and 705~\kms\ (see Figure \ref{fig:plot&spec}). Whether these minor components were real based on the observations in \citet{McCarthy+17} was not clear. However, we now are able to confirm that these components exist independently in all three of our epochs covering a time interval of more than two years. Figure \ref{relative_spectra} shows the relative difference between the spectra from each epoch of observation. No significant differences are observed, however, there is a small shift in velocity of the peak component ($\sim1$ ~\kms\ ) when comparing the 2015 August epoch and the 2017 October epoch (top panel of Figure \ref{relative_spectra}). This, and the other features of the difference spectra, is likely caused by differing array configurations between the three epochs of observation, rather than variability in the source. All strong maser emission (channels $>10\sigma$) is confined to a region smaller than approximately 1 arcsecond (projected linear size of 18pc at our assumed distance of 3.7~Mpc for NGC~4945).

We did not detect any emission from the 44.1-GHz $7_{0}~\rightarrow~6_1$A$^+$ class~I methanol maser transition (5$\sigma$ upper limit of 6 mJy at 3~km\,s$^{-1}$ spectral resolution), or the 48.4-GHz thermal methanol line (5$\sigma$ upper limit of 5 mJy at 6~km\,s$^{-1}$ spectral resolution) in our observations. Assuming an identical line profile between the two class~I maser transitions (36- and 44-GHz) we can put an upper limit on the 44/36-GHz integrated line intensity ratio of $0.07\pm0.01$. Similarly to the 2015 August epoch (described in \citet{McCarthy+17}), the 37.7-GHz class~II methanol line was not detected in either of our follow up observations (5$\sigma$ upper limit of $\sim4.5$ mJy at 3~km\,s$^{-1}$ spectral resolution). 

\subsection{Additional 7-mm spectral lines}

\subsubsection{HC$_3$N} \label{sec:hc3n}

The HC$_3$N emission (see Figure \ref{fig:hc3n and cs}) is situated within the galactic plane, along the major axis of the galactic disc, near the nucleus. Two point sources with moderately-broad emission are located south-west and north-east of the galactic core, covering velocity ranges of 410--560 \kms\ and 630--730 \kms\ respectively. These components have projected linear offsets from the nucleus of 56~pc and 110~pc for the south-west and north-east components respectively (assuming a distance of 3.7~Mpc for NGC~4945).  The emission from the south-west region seems to have multiple bright components peaking above the broad underlying pedestal, the highest of which has a flux density of 13 mJy. Conversely the north-west region appears to be simply broad emission, with a peak flux-density of 8 mJy and a strong absorption trough at $~636$~\kms\ (see Figure~\ref{fig:hc3n and cs}). This absorption feature at 636~\kms\ is observed at the same velocity as the largest absorption trough observed by \citet{Henkel+18} in their 3-mm molecular lines. We can put a $5\sigma$ upper limit of 5~mJy (at 3\kms\ spectral resolution) on the presence of HC$_3$N emission at the location of the class~I methanol maser emission. 

\subsubsection{SiO} \label{sec:sio}

\
The SiO $J = 1\rightarrow0,~\nu=0$ spectral line is most obvious in absorption towards NGC~4945. Two relatively strong absorption features are observed, one at 528~\kms\ (--8~mJy) and the other at 636~\kms\ (--9~mJy). The former is located at the 7-mm continuum peak and the latter is marginally offset (<3''; $\alpha_{2000} = 13^{\text{h}}~05^{\text{m}}~27.654^{\text{s}}, \delta_{2000}= -49^\circ~28^\prime~04.0^{\prime\prime}$) to the north-east (see top panel of Figure \ref{fig:sio}). The stronger absorption feature at 636~\kms\ is in good agreement with the feature observed by us in HC$_3$N emission, and therefore, also with the large absorption troughs seen in the 3-mm molecular lines by \citet{Henkel+18}. In addition to these two absorption sources, there is emission observed to the south-west of the nucleus, consisting of a series of narrow peaks covering the velocity range 450--540~\kms\ (lower panel of Figure \ref{fig:sio}). Compared to the other spectral lines detected in our observations these are relatively weak, with the emission not easily identifiable prior to self-calibration.

\subsubsection{CS}

Similar to HC$_3$N, CS ($J = 1 \rightarrow 0$) is also mainly arising from the galactic plane, covering the region from north-east to south-west of the galactic nucleus. The majority of emission comes from two bright regions located on either side of the nucleus within the galactic plane (see Table~\ref{tab:emission}). Both of these regions consist of broad emission covering velocity ranges of 400--550 and 640--750 \kms\ for the south-west and north-east locations, respectively (see Figure~\ref{fig:hc3n and cs}). The velocity ranges covered by the peaks of these components aligns well with those observed in the higher-frequency CS~$J =2\rightarrow1$ line towards this same source \citep{Henkel+18}. As for HC3N (Sect. 3.3.1), we identify an asymmetry between the angular offsets of the two CS $J = 1 \rightarrow 0$ peaks when compared to the 7-mm continuum peak. This offset is $4\farcs4$ and $5\farcs7$ for the SW and NE components, respectively (corresponding to a projected linear separations of 79 and 102 pc). The lack of emission observed towards the nucleus (and in the velocity range 550--640~\kms) is due to absorption. The strong absorption feature observed in HC$_3$N and SiO is also present in CS and is also seen towards the position slightly north-east of the 7-mm continuum peak (see Section \ref{sec:hc3n} and \ref{sec:sio}). This absorption is also observed in both the HCN~$J = 1 \rightarrow0$ and CS~$J =2\rightarrow1$ 3-mm lines reported by \citet{Henkel+18}.



\begin{figure*}
	\begin{minipage}[h]{\linewidth}
	\centering
	\includegraphics[scale=0.48]{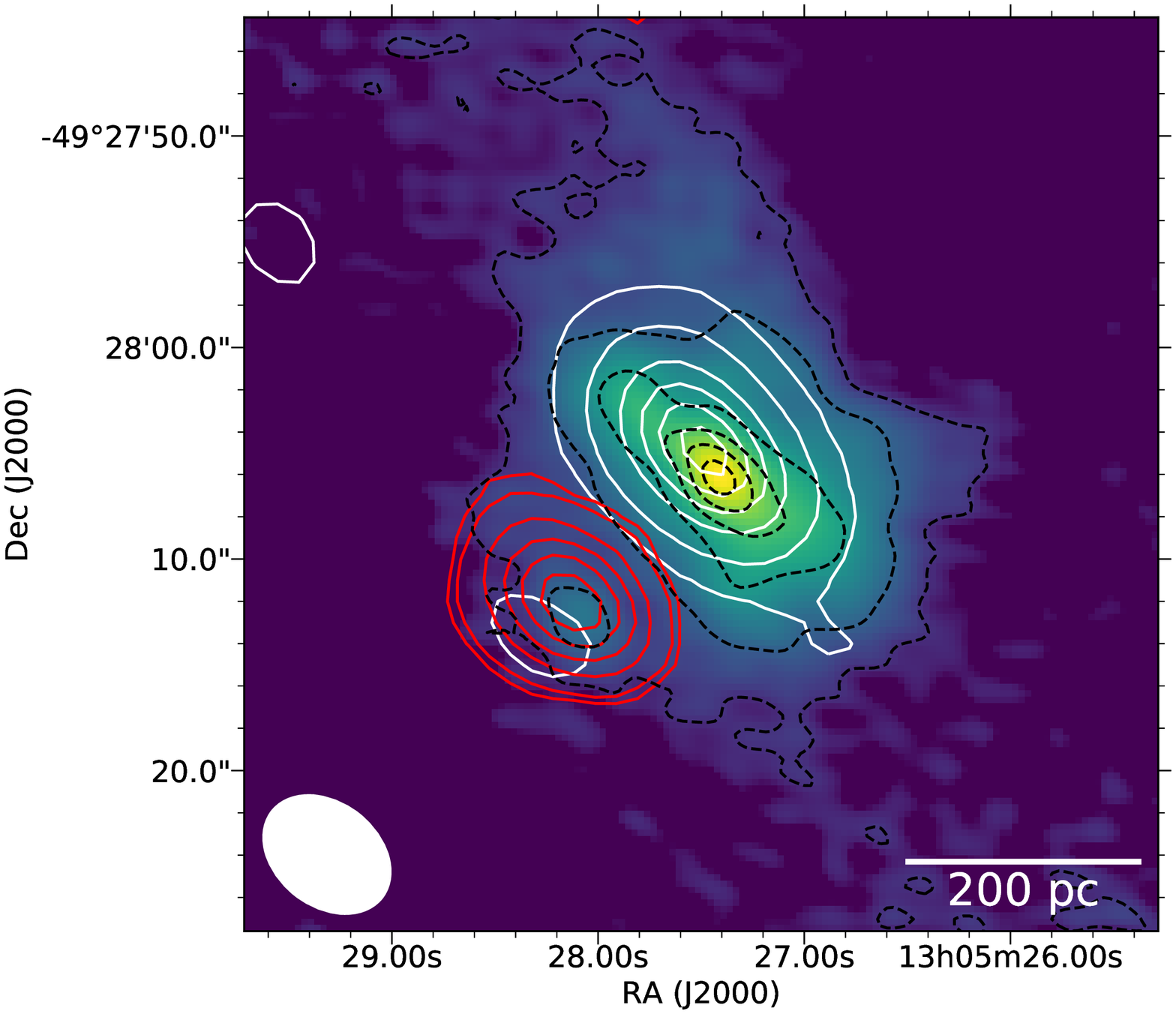}
    \end{minipage}
	\begin{minipage}[h]{0.49\linewidth}
		\centering
		\includegraphics[scale=0.7]{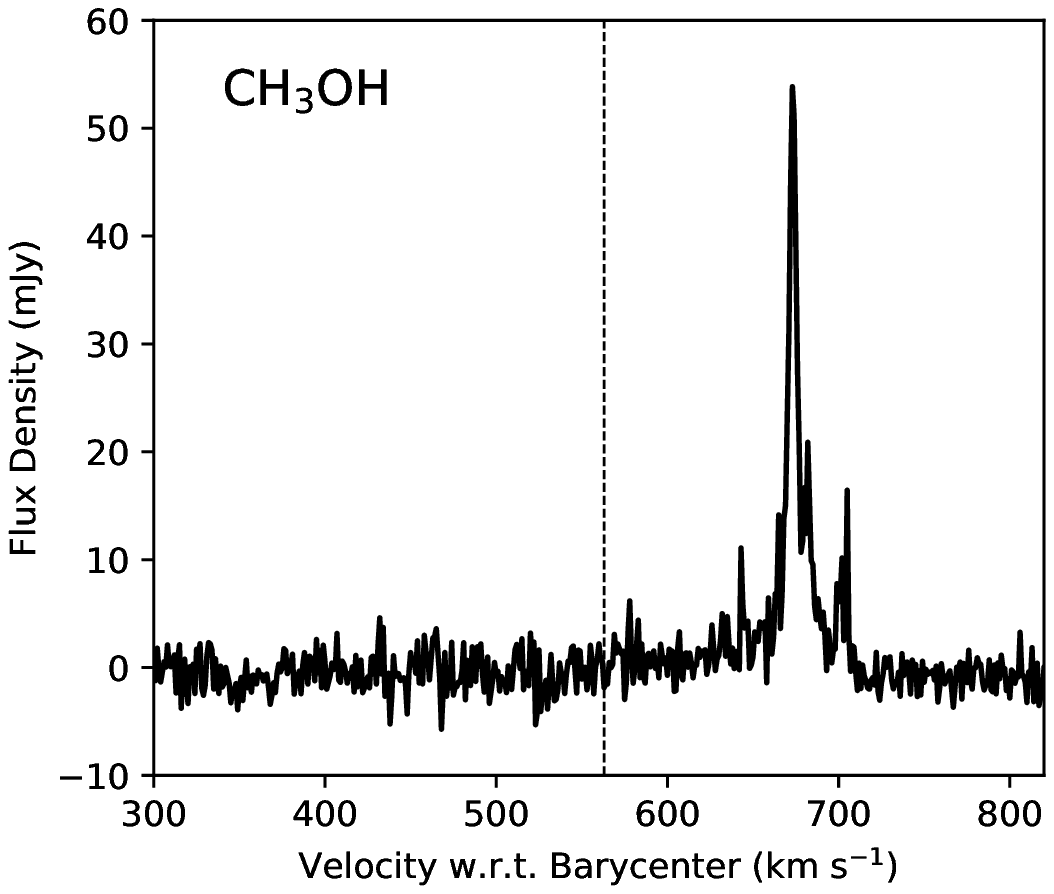}
	\end{minipage}
	\begin{minipage}[h]{0.49\linewidth}
		\centering
		\includegraphics[scale=0.70]{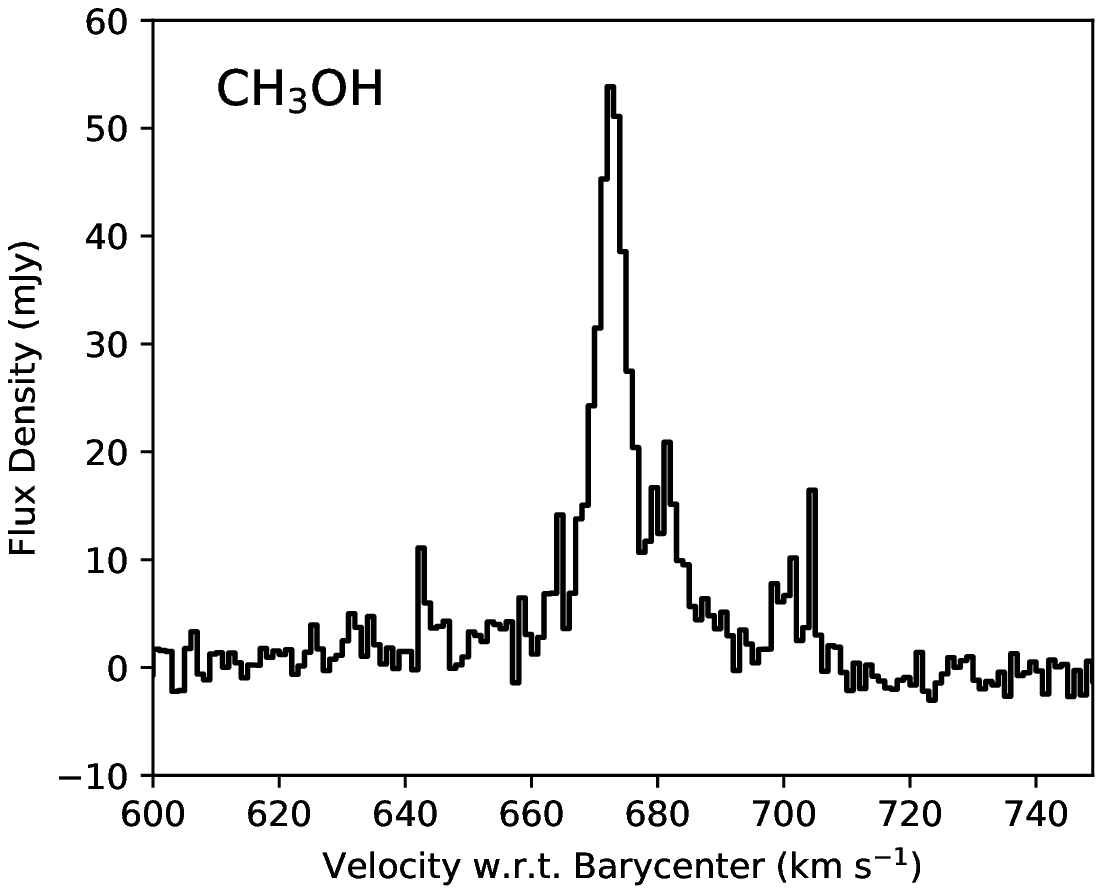}
	\end{minipage}
	\caption{Top: Integrated 36.2-GHz methanol emission (red contours 2.5\%, 10\%, 30\%, 50\%, 70\%, and 90\% of the peak of 516~mJy\,km\,s$^{-1}$\,beam$^{-1}$) and the 7-mm continuum emission (white contours 2.5\%, 10\%, 30\%, 50\%, 70\%, and 90\% of the peak of 538~mJy\,beam$^{-1}$) with  background colour map and black dashed contours of HCN $J=1 \rightarrow 0$ integrated intensity from \citet{Henkel+18} (2\%, 10\%, 30\%, 50\%, 70\%, and 90\% of the peak of 27.6 Jy\,km\,s$^{-1}$ beam$^{-1}$). Methanol and continuum emission were extracted from the combined image cube including all data from 2015 August and 2017 June/October. The white ellipse in the lower left describes the synthesised beam size for our combined observations ($6.6\times4.9$ arcseconds). Bottom Left: 36.2\,GHz spectrum from the region of peak emission within our spectral line cube (channel spacing 1\,km\,s$^{-1}$). The vertical dashed line indicates the systemic velocity of NGC\,4945 \citep{Chou+07}. Bottom Right: Same spectrum with a cropped velocity range to allow for easier differentiation between spectral features.}
	\vspace{3cm}
	\label{fig:plot&spec}
\end{figure*}

\begin{table*}
	\begin{center}
		\caption{Combined array flux-density, velocity and positional information for all molecular transitions (and continuum) observed in NGC~4945. The locations and peak flux densities given are those of the peak emission components in each transition. They were extracted using the imfit {\sc miriad} task on the spectral line cubes tabulated in Table \ref{tab:cubes}. The integrated flux-density of methanol emission tabulated here is an artifact of a minor velocity red-shift in the peak emission (between the 2017 epochs and the 2015 epoch; see Section \ref{sec:methanol_maser}). This causes a higher apparent integrated flux than observed in any individual epoch (see Table \ref{tab:individual_epoch_emission}).}
		\begin{tabular}{cccccccc} \hline
			\multicolumn{1}{c}{\bf} & \multicolumn{1}{c}{\bf Location} &\multicolumn{1}{c}{\bf RA (J2000)}  & \multicolumn{1}{c}{\bf Dec (J2000)} & \multicolumn{1}{c}{\bf $S_{pk}$} & \multicolumn{1}{c}{\bf $S$} & \multicolumn{1}{c}{\bf $V_{pk}$} & \multicolumn{1}{c}{\bf $V_{Range}$}  \\
			& & \multicolumn{1}{c}{\bf $h$~~~$m$~~~$s$}& \multicolumn{1}{c}{\bf $^\circ$~~~$\prime$~~~$\prime\prime$} & (mJy) & (mJy\,km\,s$^{-1}$) & (km\,s$^{-1}$) & (km\,s$^{-1}$) \\  \hline
			Methanol & SE &  13 05 28.167 & $-$49 28 12.1 & $55\pm6$ & $516\pm10$  & 673 & 660 -- 720  \\    
			HC$_3$N & SW & 13 05 26.936 & $-$49 28 07.9 & $13.5\pm3$ & $399\pm14$  & 479 & 390 -- 550 \\
			& NE & 13 05 27.621 & $-$49 28 02.1 & $8.4\pm3$ & $177\pm14$ & 686 & 550 -- 720 \\  
			CS & NE & 13 05 27.961 & $-$49 28 01.9 & $63.0\pm10$ & $2692\pm300$  & 705 & 640 -- 750 \\ 
			 & SW & 13 05 27.196 & $-$49 28 08.3 & $46.2\pm10$ & $3346\pm300$ & 465 & 400 -- 550 \\ 
			Continuum & - & 13 05 27.467 & $-$49 28 04.8 & $295\pm15$ & $436\pm33$  &  &  \\   \hline
		\end{tabular} \label{tab:emission}		
	\end{center}
\end{table*}

\section{Discussion}

\subsection{The masing methanol of NGC~4945} \label{sec:methanol_maser}

Follow up observations of the $4_{-1}\rightarrow3_0E$ methanol transition in NGC~4945 have allowed for higher fidelity imaging of the masing region. The main advantage with respect to the data previously presented by \citet{McCarthy+17} is the significantly improved uv-coverage. We find good agreement (to within an arcsecond) between the previously reported class~I maser location ($\alpha_{2000} = 13^{\text{h}}~05^{\text{m}}~28.093^{\text{s}}, \delta_{2000}= -49^\circ~28^\prime~12.3^{\prime\prime}$) and the position determined using all three epochs of observation ($\alpha_{2000} = 13^{\text{h}}~05^{\text{m}}~28.167^{\text{s}}, \delta_{2000}= -49^\circ~28^\prime~12.1^{\prime\prime}$; see Table \ref{tab:emission}), with the standard deviation in our fit more than an order of magnitude lower ($\sim0.02^{\prime\prime}$ compared to the previous $\sim0.4^{\prime\prime}$). The astrometric accuracy to which we can determine the maser position is limited by the systematics of our observing configuration (array configuration etc.), rather than by errors in our position fitting.


When the initial detection of 36.2-GHz maser emission in NGC~4945 was reported in \citet{McCarthy+17}, its galactic bar was thought to have a position angle of 33$^{\text{o}}$ with an azimuth angle with respect to the plane of the galaxy of 40$^{\text{o}}$ based on CO and HI observations by \citet{Ott+01}. Recent investigations by \citet{Henkel+18} have concluded that the bar dynamics are drastically different than previously thought, with the bar elongated approximately east-west in the plane of the sky. \citeauthor{Henkel+18} suggest that the class~I methanol maser is likely associated with the front side of the bar or the south-eastern spiral arm, also being part of the front-side of the galaxy. Assuming the maser emission is exactly in the plane of the galactic disk and that the inclination of NGC~4945 is approximately 75$^{\text{o}}$, the class~I maser would have a galacto-centric radius of $\sim650$~pc (note that if you consider a disk with some thickness, a range of valid radii about this value should be considered). This puts it outside of the estimated 300~pc maximum radius that the bar extends, indicating the masing region is likely instead located within the inner south-eastern spiral arm. This is consistent with what is observed towards other extragalactic class~I methanol maser sources, such as NGC~253 and IC~342, where the class~I masers are offset from their dynamical centre, but close to the interfacing regions of the galactic bar \citep{Ellingsen+17b,Gorski+18}. If the masing methanol is located in this interface environment, it is consistent with the hypothesis that these masing regions result from large-scale low-velocity shocks \citep{Ellingsen+14,Ellingsen+17b, Gorski+18}.

The methanol maser emission appears to be spatially coincident with the south-eastern hotspot (see HCN colour map in Figure \ref{fig:plot&spec}) observed in the HCN and CS (J~=~$2\rightarrow1$) dense gas tracers by \citet{Henkel+18}. This hotspot is also nearby the location of `Knot B' a structure observed previously in both H$\alpha$ and Paschen-alpha \citep{Moorwood+96, Marconi+00}. \citet[][their Sect. 4.1.3]{Henkel+18} find a small offset between Knot B and this secondary molecular peak, i.e. the secondary molecular peak is offset by (+3$^{\prime\prime}$,-2$^{\prime\prime}$) from knot B. This corresponds to about 3$\farcs$5 or about 70 pc. They speculate that this hotspot is a giant molecular cloud complex and that Knot B is the star-forming front side of this region. We find the 36.2-GHz methanol maser emission is located closer (offset by $\sim1^{\prime\prime}$) to the molecular peak, rather than the region of star-formation represented by Knot B. The peak velocity of the class~I maser emission is red-shifted relative to the velocity of this hotspot ($\sim660$~\kms), however, only by approximately 15~\kms. Interestingly, we do not see any sign of a CS hotspot in the lower frequency CS transition that we have observed (see Section \ref{sec:CS}). This is not entirely unexpected, as the feature is not strong in the $J = 2 \rightarrow 1$ transition and may, due to lower statistical weights, be even weaker in the $J = 1 \rightarrow 0$ line we observe here.


In Galactic sources, 36.2- and 44.1-GHz masers are often observed together, with the 44.1-GHz transition generally being brighter \citep{Voronkov+14}. The lack of observed 44.1-GHz emission towards NGC~4945, alongside the narrowness of the observed emission, reinforces the conclusion that the class~I maser in NGC~4945 is not due to a large number of Galactic-like star formation regions in a confined area. 


\begin{figure}
	\captionsetup[subfigure]{justification=centering}
	\begin{center}
		\begin{minipage}[t]{1.0\linewidth}
			\begin{center}
				\subfloat[October 2017 -- August 2015 ]{\includegraphics[scale=0.63]{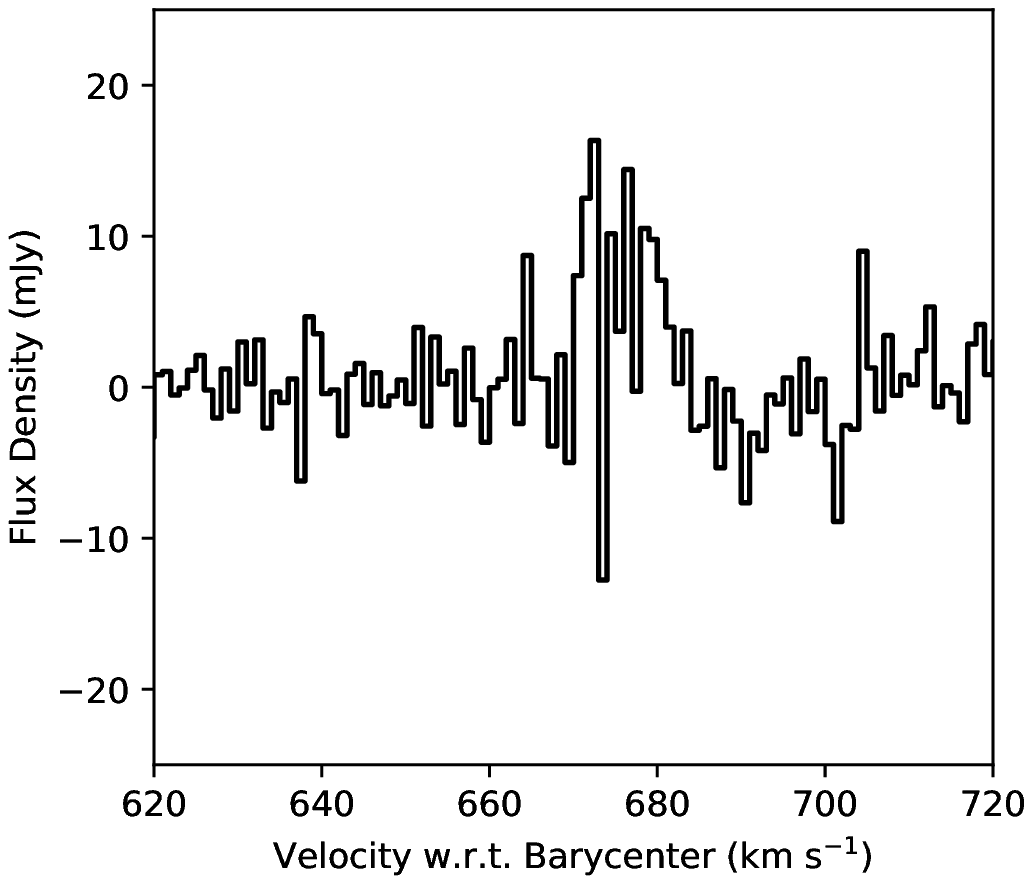}} \
				\subfloat[October 2017 -- June 2017]{\includegraphics[scale=0.63]{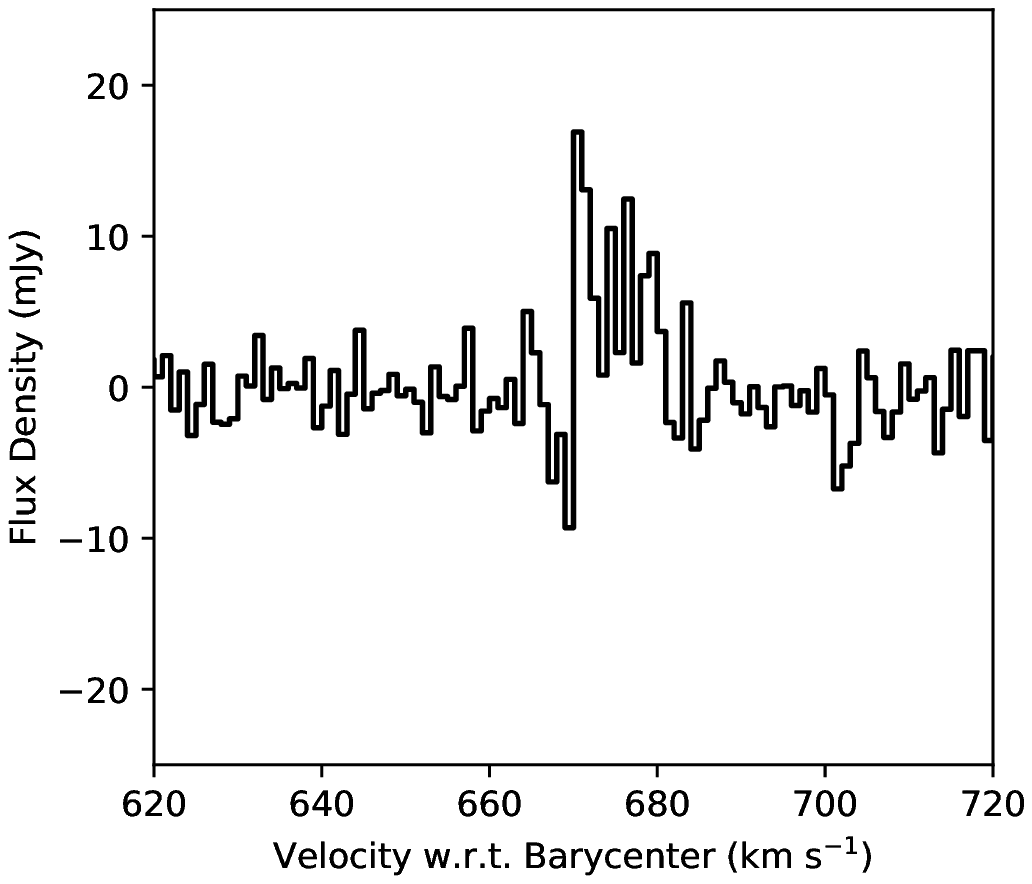}} \
			\end{center}
		\end{minipage}
	\end{center}
	\caption{Differences between 36.2-GHz methanol maser spectra between individual epochs. Both 2015 August and 2017 June were subtracted from the 2017 October spectrum with the top panel (a) corresponding to the former case and bottom panel (b) to the latter. Vertical scales for each panel are identical for ease of comparison. Emission from the 2017 October epoch appears broader than either of the other epochs of observations (likely due to the more compact array configuration). Additionally, the 1~\kms\ peak shift can be seen in panel (a).}
	\vspace{3cm}
    \label{relative_spectra}
\end{figure}


\begin{table}
	\begin{center}
		\caption{36.2\,GHz methanol and continuum emission toward NGC\,4945 from individual observation epochs.}
		\begin{tabular}{ccccc} \hline
			\multicolumn{1}{c}{\bf} & \multicolumn{1}{c}{\bf Epoch} & \multicolumn{1}{c}{\bf Array} & \multicolumn{1}{c}{\bf $S_{pk}$} & \multicolumn{1}{c}{\bf $S$}   \\
			& & & (mJy) & (mJy\,km\,s$^{-1}$)  \\  \hline
			Methanol & Aug2015 & EW367 & 49 & $294\pm23$   \\ 
			& Jun2017 & H214 & 48 & $306\pm40$   \\ 
			& Oct2017 & H168  & 54  & $339\pm18$     \\ 
			Continuum & Aug2015 & EW367 & 313 &  $414\pm40$      \\
			& Jun2017 & H214 & 236 & $348\pm45$  \\
			& Oct2017 & H168 & 385 & $489\pm18$   \\   \hline
		\end{tabular} \label{tab:individual_epoch_emission}		
	\end{center}
\end{table}

As our observations were made over three independent epochs, this gives us the chance to compare and contrast the properties of the observed 36.2-GHz emission from NGC~4945 at different times and array configurations. The peak and integrated flux densities from each epoch are tabulated in Table \ref{tab:individual_epoch_emission} for the 36.2-GHz emission zoom band (both masing and continuum). The flux-density of the maser emission across all three epochs is consistent with what we expect from a compact region, with the difference in peak and integrated flux density between the least and most compact array configurations (EW352 and H168 respectively) only varying by approximately 10\%. Comparison of the continuum emission across the different array configurations is not as straight forward, with the 2017 June epoch showing a significant drop in flux-density (both integrated and peak) compared to the other two epochs (see Table \ref{tab:individual_epoch_emission}). This same decrease in flux-density is not observed in any of the calibrator sources for this epoch, with only the continuum emission of NGC~4945 lower than expected. Higher levels of decorrelation due to sub-optimal observing conditions may be the cause of this, as this can affect continuum and line emission differently.

The secondary 36.2-GHz component (red-shifted with respect to the peak) mentioned in \citet{McCarthy+17} has been observed independently in all three epochs, increasing confidence that it is a real component of maser emission (7$\sigma$ in combined epoch data). When combining data from all three epochs, we can discern additional features in the 36.2-GHz spectrum. Firstly, another low flux-density component (5$\sigma$ detection in combined epoch data) appears blue-shifted in relation to the peak emission. Interestingly, these two minor components, are roughly symmetrically distributed about the central peak emission. Additionally, in the combined data set, the central peak appears to split into two components of maser emission with a small relative velocity offset. This symmetric distribution of masing emission is suggestive of the spectra seen towards edge on disk structures, similar to those observed from H$_2$O megamasers towards the nucleus, with the bright emission at the systemic velocity and the red- and blue-shifted components at the approaching and receding edges of the disk. Assuming an edge on disk, we identify a rotational velocity of $V_{rot} = 30$~\kms\ and a velocity drift of $0.5$~\kms\,yr$^{-1}$ (obtained from peak shift discussed in Section \ref{RES:Methanol}). These values correspond to a disk with radius of $\sim380$~AU with enclosed mass of $\sim390$~M$_{\sun}$ \citep{Ishihara+01}. However, this is highly speculative as our resolution is not high-enough to satisfactorily determine whether the positions of these individual components show any ordered structure.


\subsection{CS: comparison with 3-mm observations}\label{sec:CS}

With the recent addition of ALMA to the southern hemisphere's astrophysical toolbox, the dynamics and molecular structure of the central region of NGC~4945 are now beginning to be more clearly understood \citep[e.g.][]{Bendo+16, Henkel+18}. Our observations of the CS $J=1\rightarrow0$ emission towards NGC~4945 show strong agreement, in terms of position and velocity of the bright regions north-east and south-west of the galactic nucleus, when compared to the CS and HCN emission reported by \citet{Henkel+18}. We identify a line ratio of approximately 4 between the CS $J = 2\rightarrow 1$ and CS $J=1\rightarrow0$ emission \citep{Henkel+18}. This indicates the the CS $J=1\rightarrow0$ emission is optically thin and highly excited. As mentioned in Section \ref{sec:methanol_maser}, we see no evidence for the south-eastern hotspot in the CS $J = 1 \rightarrow 0$ transition (5$\sigma$ upper limit of 15 mJy in a 3~\kms\ channel), likely due to the large discrepancy in flux densities between the two CS transitions. Possibly, longer on-source integration times would reveal some weak emission from this location. It should also be noted the synthesised beam of our observations is significantly larger than that of the ALMA observations presented by \citeauthor{Henkel+18} ($5.2 \times 3.3$ compared to  $2.6\times1.4$ arcseconds).

\citet{Henkel+18} propose that the east-west oriented bar in NGC~4945 can be directly seen in their CS $J = 2\rightarrow1$ and HCN $J = 1\rightarrow 0$ lines, when integrating over that narrow velocity range near the systemic velocity of the galaxy, which is least affected by absorption. A direct connection can be seen in their HCN integrated intensity map, and the same east-west structure is even more clearly seen in the CS emission. Comparing our lower frequency CS $J=1\rightarrow0$ emission across this velocity range (585 -- 612~\kms), we are also able to clearly see this bar structure (see Figure \ref{fig:cs_bar_emission}), though not as extended as evident in the higher frequency transitions.


\begin{figure*}
	\begin{minipage}[h]{0.49\linewidth}
		\centering
		\includegraphics[scale=0.65]{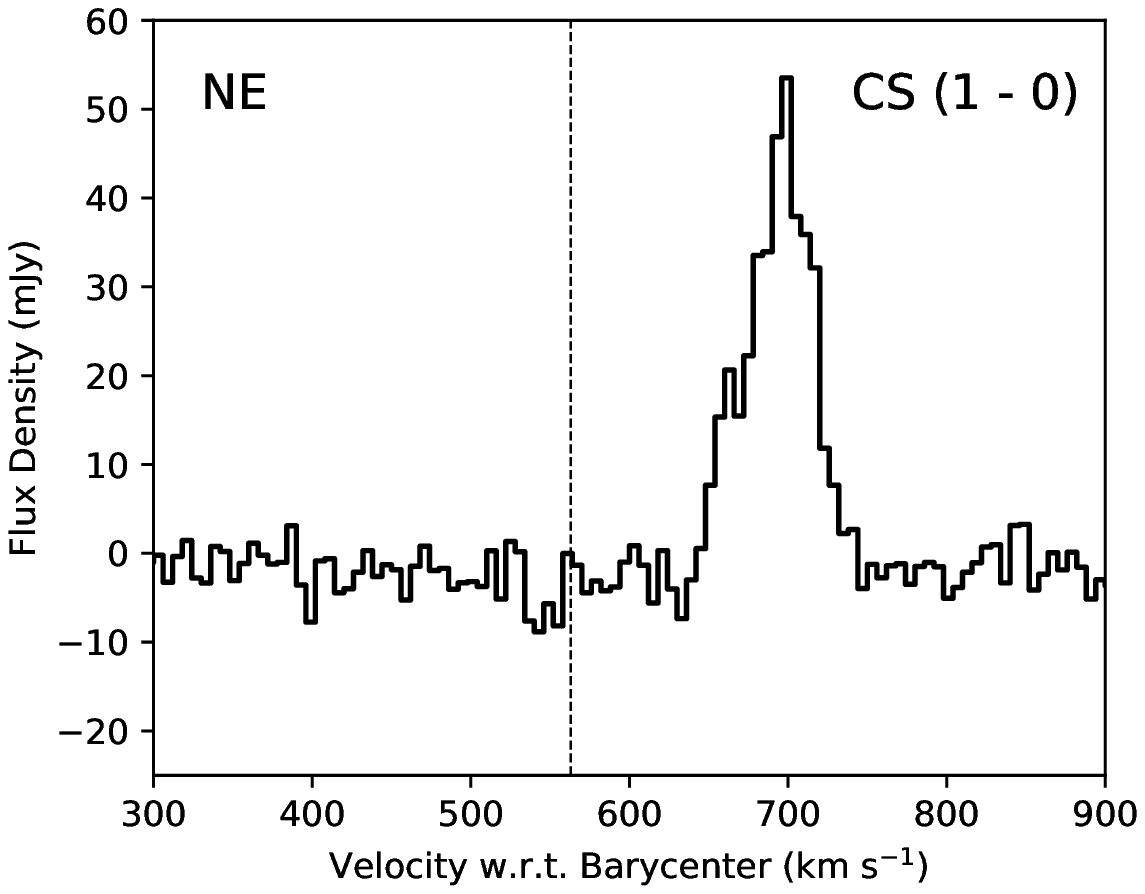}
	\end{minipage}
	\begin{minipage}[h]{0.49\linewidth}
		\centering
		\includegraphics[scale=0.65]{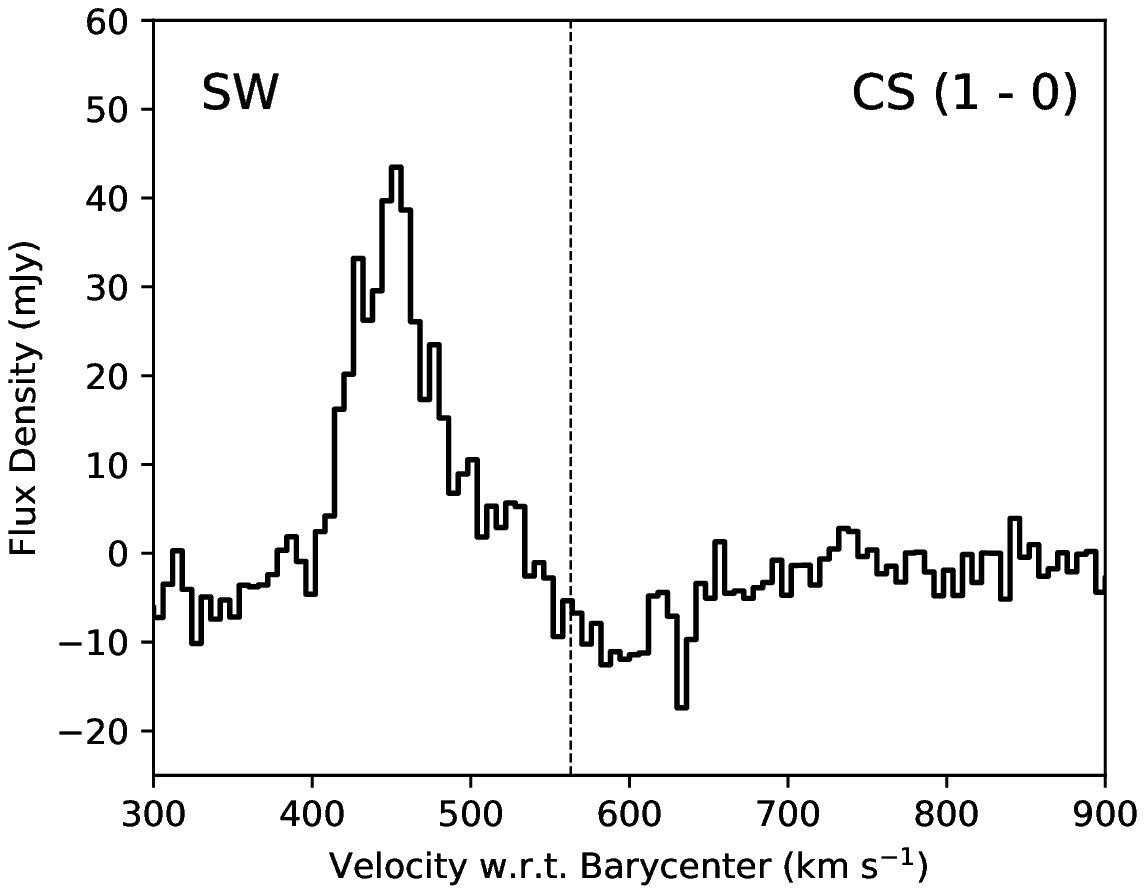}
	\end{minipage}
	\begin{minipage}[h]{\linewidth}
		\centering
		\includegraphics[scale=0.45]{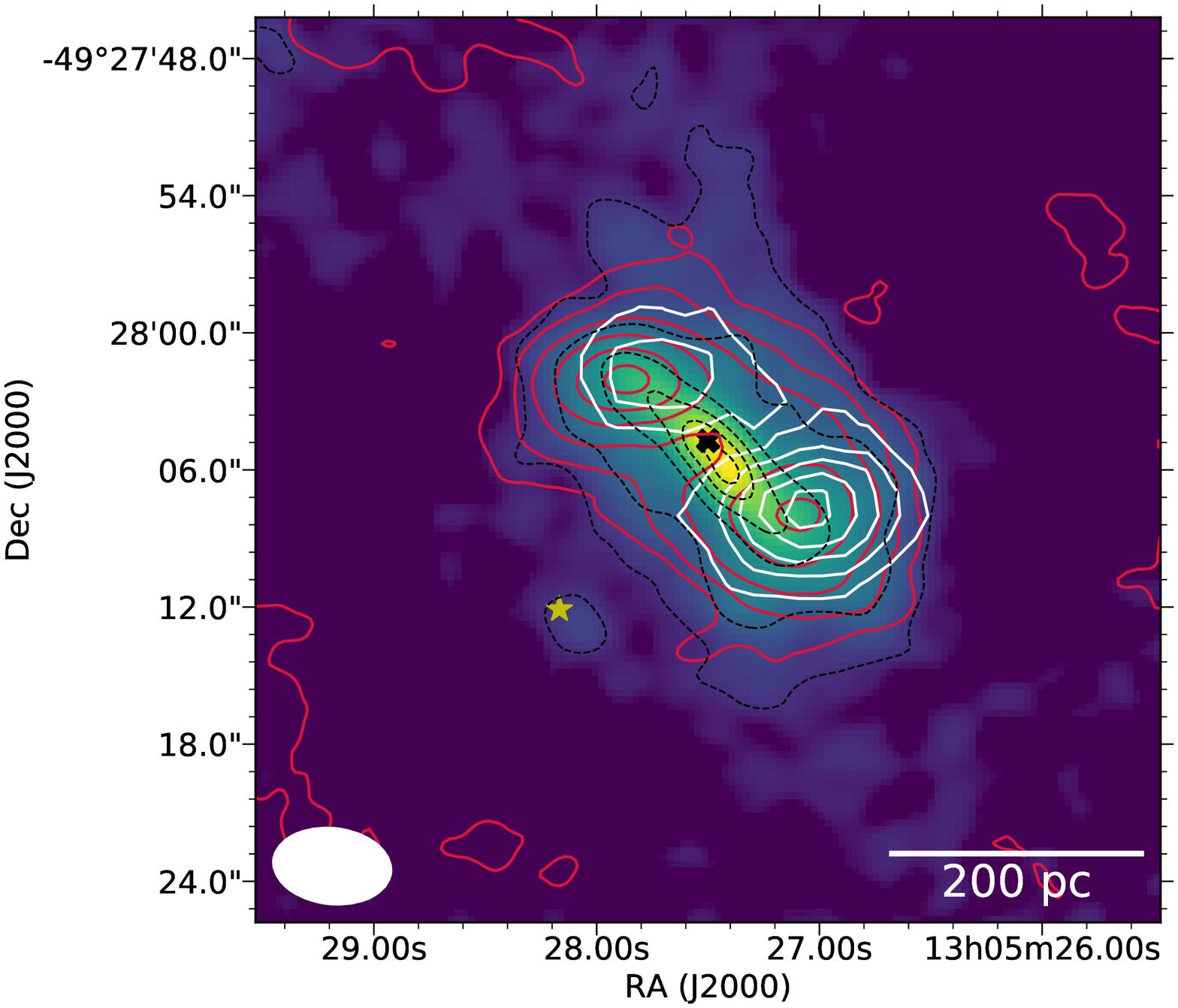}
	\end{minipage}
	\begin{minipage}[h]{0.49\linewidth}
		\centering
		\includegraphics[scale=0.65]{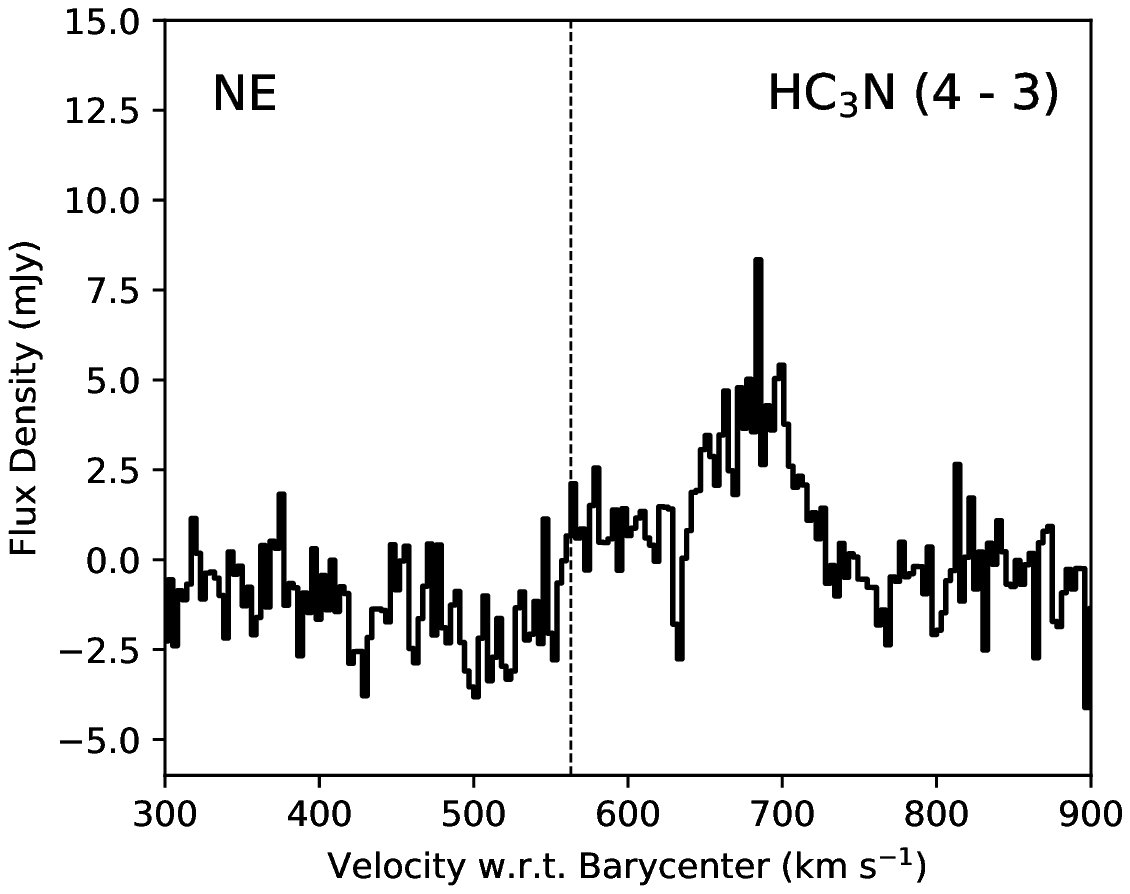}
	\end{minipage}
	\begin{minipage}[h]{0.49\linewidth}
		\centering
		\includegraphics[scale=0.65]{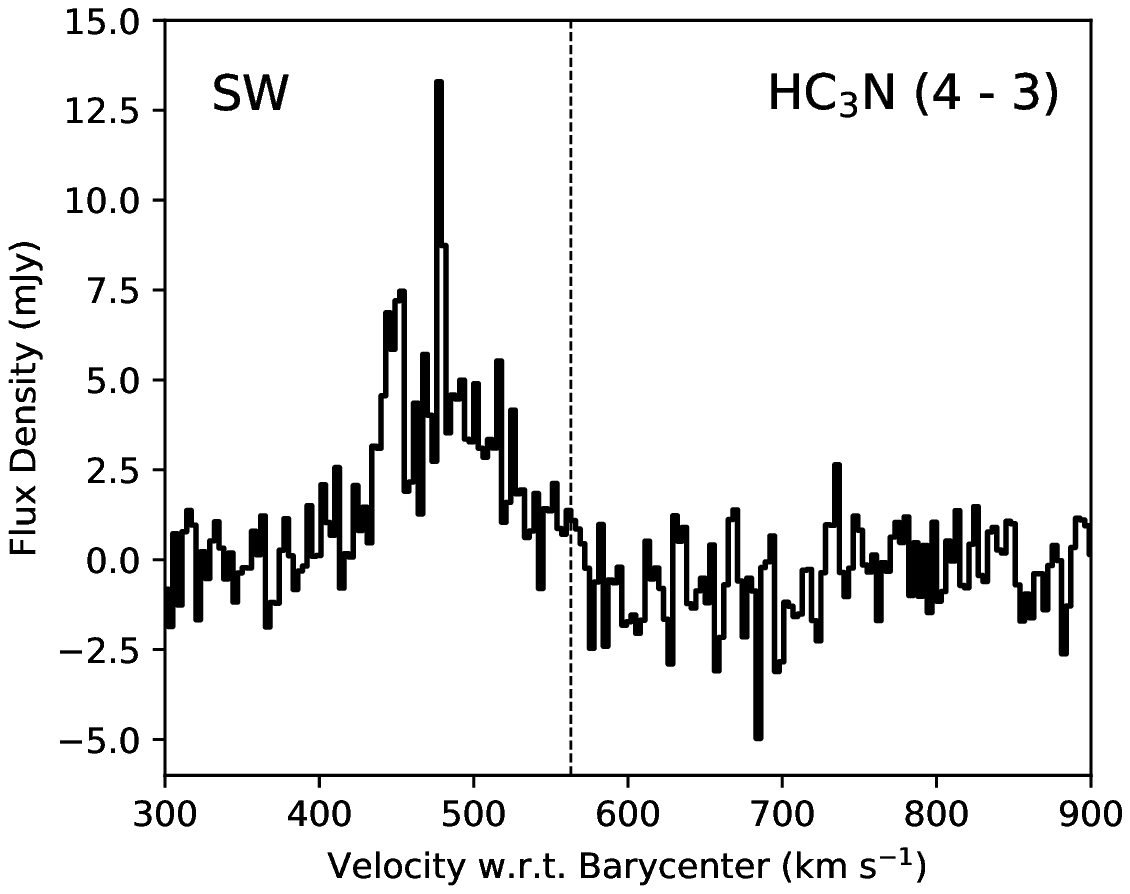}
	\end{minipage}
	\caption{Centre: 36.4-GHz HC$_3$N integrated emission (white contours 10\%, 30\%, 50\%, 70\%, and 90\% of the 399 mJy\,km\,s$^{-1}$ beam$^{-1}$ peak, restored beam: $6.3\times4.8$ arcseconds) and CS $J=1\rightarrow0$ emission (red contours 10\%, 30\%, 50\%, 70\%, and 90\% of the 2130 mJy\,km\,s$^{-1}$ beam$^{-1}$ peak, restored beam: $5.2\times3.3$ arcseconds), overlaid on a colour map (with accompanying dashed black contours) of the CS $J=2 \rightarrow 1$ integrated intensity from \citet{Henkel+18} (contour levels 2\%, 10\%, 30\%, 50\%, 70\%, and 90\% of the peak of the 11.5 Jy\,km\,s$^{-1}$ beam$^{-1}$). The black cross and yellow star indicate the position of the 7-mm continuum peak and 36.2-GHz maser location respectively. The white ellipse represents the synthesized beam of the CS $J = 1\rightarrow0$ observations. Top: Spectra of the north-east and south-west CS $J=1\rightarrow0$ components (from left to right) extracted at the locations of peak emission from the image cube with a spectral resolution of 6~\kms. Bottom: Spectra of the north-east and south-west HC$_3$N $J=4\rightarrow3$ components (from left to right) extracted at the location of peak emission from the image cube with a spectral resolution of 3~\kms.}
	\vspace{3cm}
	\label{fig:hc3n and cs}
\end{figure*}

\begin{figure}
	\captionsetup[subfigure]{justification=centering}
	\begin{center}
		\begin{minipage}[t]{1.0\linewidth}
			\begin{center}
				\subfloat[SiO absorption]{\includegraphics[scale=0.63]{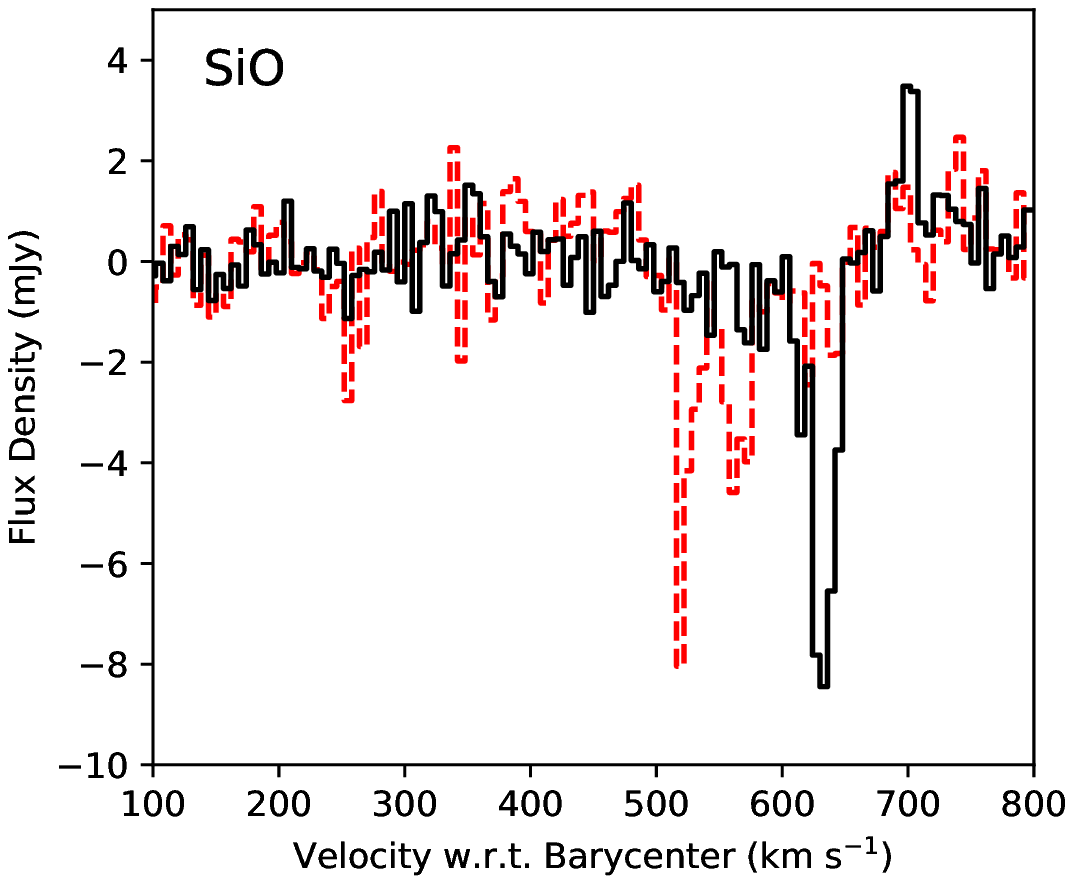}} \
				\subfloat[SiO emission]{\includegraphics[scale=0.63]{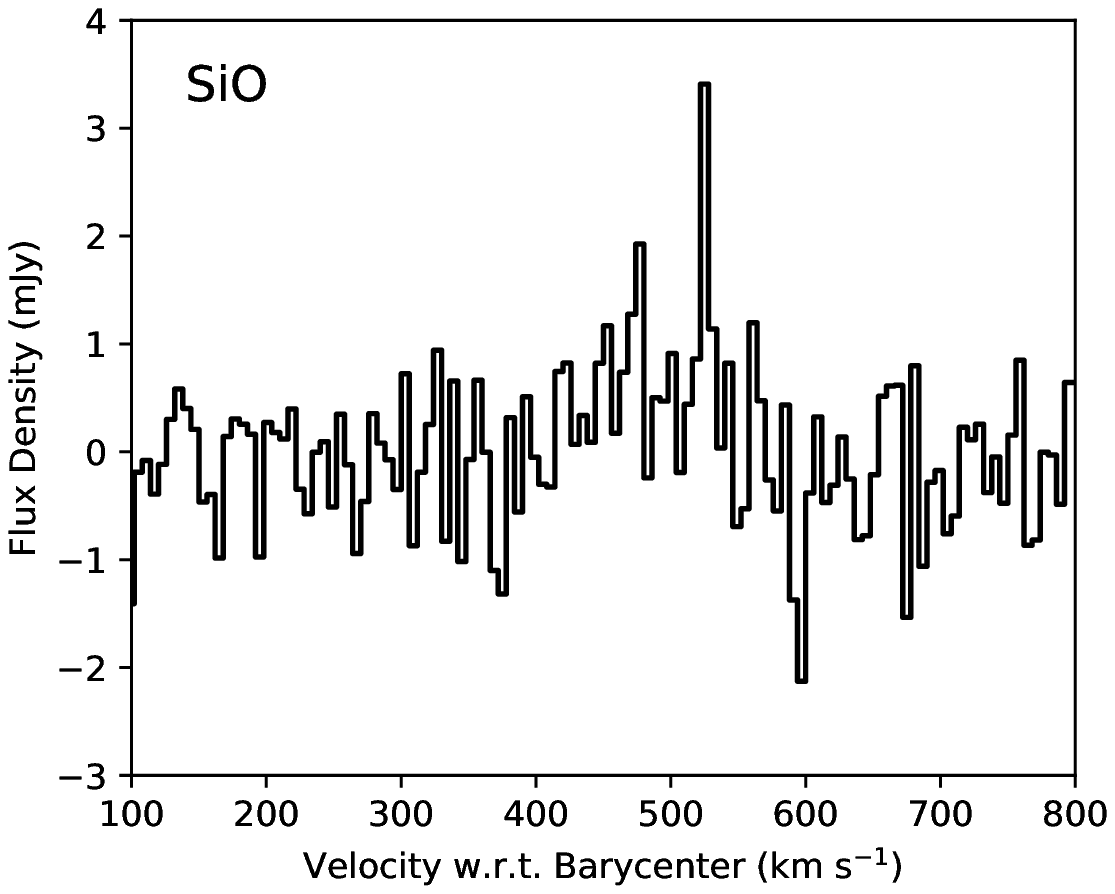}} \
			\end{center}
		\end{minipage}
	\end{center}
	\caption{Top: SiO absorption spectra for the two main components in NGC~4945. The red-dashed spectrum represents absorption towards the 7-mm continuum peak (see Table \ref{tab:emission}). The black spectrum represents absorption from the component offset 3$^{\prime\prime}$ to the north-east of the continuum peak. Both spectra have a spectral resolution of 6~\kms. Bottom: Spectrum of weak SiO emission from south-west of the continuum peak taken from a spectral line cube with spectral resolution of 6~\kms.}
	\vspace{3cm}
	\label{fig:sio}
\end{figure}
\subsection{Nature of HC$_3$N emission in NGC~4945}\label{sec:hc3n_emission}

Determining the nature of the HC$_3$N $J = 4 \rightarrow 3$ emission in NGC~4945 is difficult as there are not many known Galactic or extragalactic sources displaying maser emission from this species. The Sgr B2 complex has been observed in various HC$_3$N transitions, with maser emission detected in the $J = 1 \rightarrow 0$ transition \citep{Hunt+99}. \citet{McGee+77} report $J = 4\rightarrow3$ emission in Sgr B2 with an integrated flux density of $\sim660$~Jy~\kms. Comparing this flux density to the total integrated intensity across both the north-east and south-west components in NGC~4945, we see that NGC~4945 is approximately a factor of 190 times more luminous than Sgr~B2 \citep[assuming a distance of 7.9~kpc for Sgr~B2 and 3.7-Mpc for NGC~4945;][]{Reid+09, Tully+13}. Comparing the individual components to the HC$_3$N $J=4\rightarrow3$ emission in Sgr~B2, we see that the south-eastern HC$_3$N component in NGC~4945 is $\sim130$ times more luminous. 

Recently, HC$_3$N $J = 4 \rightarrow 3$ emission towards NGC~253 has been reported to be the result of maser processes \citep{Ellingsen+17}. The emission we observe towards NGC~4945 bares some similarities to that seen in NGC~253, especially when comparing it on a component to component scale. The total integrated flux of the HC$_3$N~$J = 4\rightarrow3$ emission in NGC4945 is almost a factor of 10 lower than that observed in NGC~253 \citep{Ellingsen+17}. However, when looking at the individual HC$_3$N components (2 in NGC~4945, 7 in NGC~253) across both sources, we see that the integrated flux densities of individual components are very similar. The most luminous component observed towards NGC~253 is 443~mJy~\kms\ \citep{Ellingsen+17}. This would correspond to a flux-density of 374~mJy~\kms\ in NGC~4945 \citep[assuming a distance of 3.4~Mpc for NGC~253 and 3.7~Mpc for NGC4945;][]{Dalcanton+09, Tully+13}, very similar to the highest flux-density HC$_3$N component (the south-western one) observed towards NGC~4945. Likewise, the north-eastern HC$_3$N component in NGC~4945 also shows a flux-density which is similar to the individual spots observed in NGC~253. It is important to note that the flux-density values reported for NGC~253 in \citet{Ellingsen+17} are using the combined EW367 and 1.5A array configurations, both of which have longer maximum baselines than the two array configurations used for our observations (H168 and H214). Therefore, missing flux from extended emission makes comparison of flux-densities unreliable.

The two regions of HC$_3$N emission in NGC~4945 have close spatial correlations with the observed CS $J = 1\rightarrow 0$ emission (see Figure~\ref{fig:hc3n and cs}). Emission from the two HC$_3$N point sources appears to heavily overlap with the most luminous regions of CS north-east and south-west of the core, with similar velocities for peak emission (see Table \ref{tab:emission}). This is similar to what is seen in NGC~253, where the HC$_3$N spots correlate with the position of molecular clouds \citep{Leroy+15,Ellingsen+17}. These molecular clouds are defined by regions of excess emission from dense gas tracers such as CS and HCN \citep{Meier+15, Leroy+15}. The linear offsets from the nucleus (110 and 56~pc towards the south-west and north-east respectively) for the HC$_3$N spots in NGC~4945, puts them close to the radius of the outer nuclear disk and inner bar interface \citep{Henkel+18}. 

\citet{Ellingsen+17} concluded that at least one of the HC$_3$N spots they observed in NGC~4945 was due to a maser process. However, it is difficult to evaluate whether this is also the case for either region observed towards NGC~4945. This is partly because of the limited angular resolution of our observations, which prevents us from determining how compact these regions are.  Unlike the actively studied maser species such as methanol and H$_2$O, the maser process of HC$_3$N is much less clear. Therefore, we do not know whether a typical HC$_3$N maser forms in a compact region, as observed in those other molecular species. If we assume it does, based on the maser emission detected towards NGC~253, then a follow-up of the HC$_3$N observations in NGC~4945 at higher angular resolution will help us determine the nature of these regions. Additionally, detailed investigation into the nuclear molecular clouds has not been conducted in NGC~4945. Information about the environments housing these spots would provide useful insight into the processes governing the HC$_3$N emission.
\\
\begin{figure}
	\includegraphics[width=\linewidth]{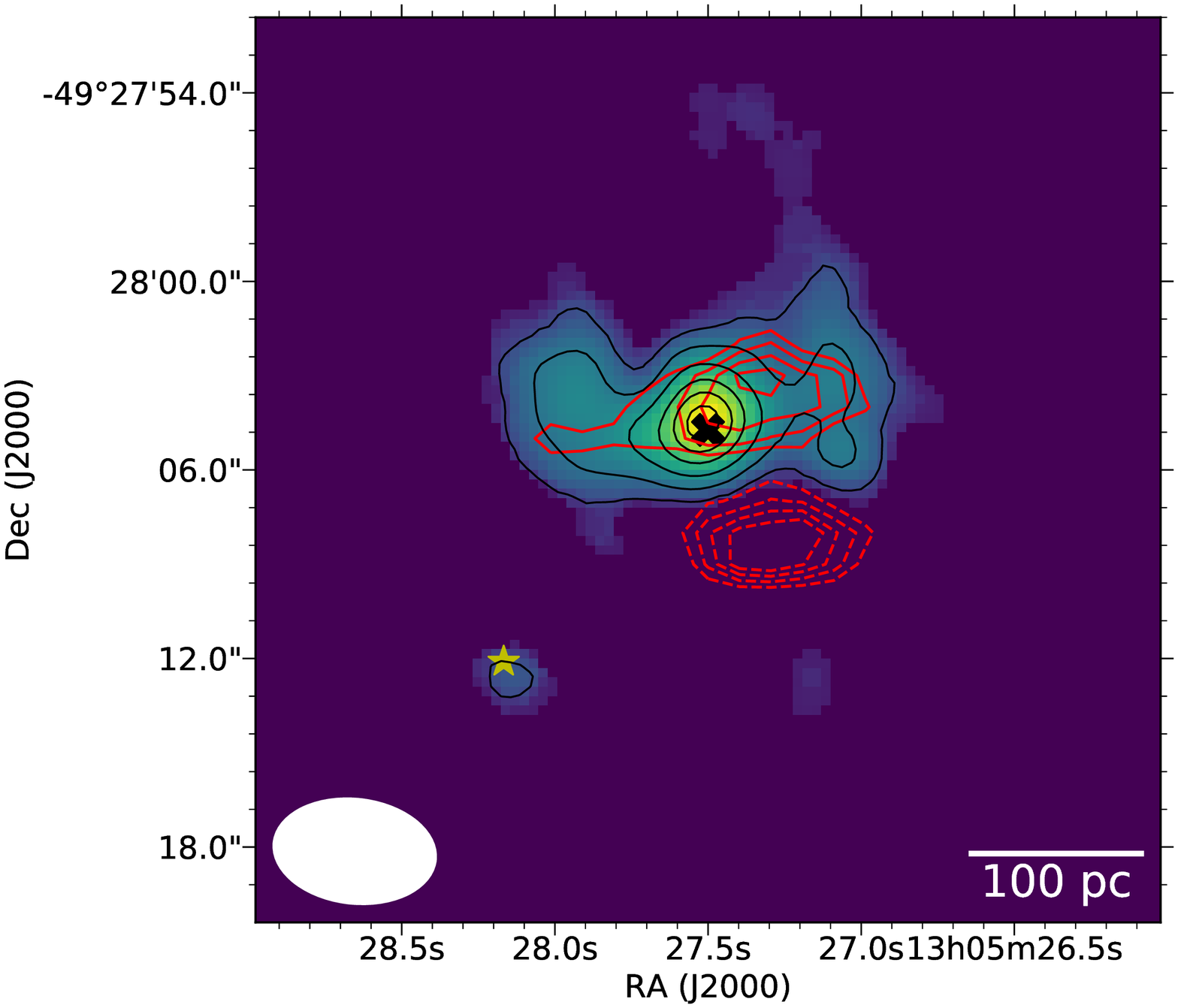}
	\caption{Integrated intensity of 49-GHz CS $J=1\rightarrow0$ emission over the velocity range 585 -- 612~\kms\ (red contours 30\%, 50\%, 70\%, and 90\% of the 187 mJy\,km\,s$^{-1}$ beam$^{-1}$ peak, dashed contours represent the same levels for absorption). This is the same velocity range where the east-west structure of the bar is visible in the CS $J=2\rightarrow1$ emission presented by \citet{Henkel+18}. The background colour map represents CS $J=2 \rightarrow 1$ integrated intensity across this same reduced velocity range \citep[black contours 30\%, 50\%, 70\%, and 90\% of the 2.4 Jy\,km\,s$^{-1}$ beam$^{-1}$ peak;][]{Henkel+18}. The black cross and yellow star indicate the position of the 7-mm continuum peak and 36.2-GHz maser location respectively.}
	\label{fig:cs_bar_emission}
\end{figure} 

\subsection{Comparison of the CH$_3$OH masing environment in NGC~4945 to other relevant sources}

In the absence of direct information regarding the physical conditions responsible for these extragalactic masers, comparisons with similar sources may reveal useful properties of the masing regions. These comparisons can both help understand the phenomenon and allow for future searches with better constrained source samples.

This section will first compare the masing region of NGC~4945 to those observed within NGC~253, and subsequently compare both of these sources to the Galactic giant molecular cloud G\,1.6-0.025.

\subsubsection{NGC~253}

The sample size of known extragalactic class~I methanol masers is currently very small. This makes the nature of such emission difficult to discern. The best match up of host-galaxies from the available sample is between NGC~253 and NGC~4945. Similarly to NGC~4945, NGC~253 is a nearby barred spiral starburst galaxy, with recently detected class~I methanol maser emission at 36.2-GHz \citep{Sakamoto+11, Ellingsen+14}. These two sources are similarly distant \citep[3.4~Mpc for NGC~253 and 3.7~Mpc for NGC~4945;][]{Dalcanton+09, Tully+13}, and have methanol maser emission of comparable luminosity. Due to these factors, these two sources are the best candidates for a direct comparison of environments hosting the masing methanol.

Our follow up observations have allowed us to more accurately compare the integrated intensity of NGC~4945 to NGC~253, as both sources have been observed with the same array configuration (H168) and are at similar distances. We find that the single region of class~I maser emission in NGC~4945 has an integrated flux-density 50\% greater than either emission region in NGC~253 \citep{Ellingsen+14, Ellingsen+17b}. In addition to being brighter, the 36.2-GHz methanol maser component towards NGC~4945 has a much narrower linewidth, $\sim$8~\kms\, compared to those observed towards any of the regions in NGC~253 \citep[$\ga20$~\kms\ at comparable angular and spectral resolution; ][]{Ellingsen+17b}.

There is a strong correlation between the masing regions of NGC~253 and the presence of large molecular clouds \citep{Leroy+15, Ellingsen+17b}. As mentioned in Section \ref{sec:methanol_maser}, images presented by \citet{Henkel+18} show the methanol masing region in NGC~4945 is projected onto a HCN and CS hotspot. This hotspot may mark the location of a molecular cloud complex, similar to those identified in NGC~253 \citep{Leroy+15}. 36.4-GHz HC$_3$N emission is observed towards all of the class I methanol maser emission regions in NGC~253 \citep{Ellingsen+17}. This same relationship is not observed towards NGC~4945, where we place a 5$\sigma$ upper limit of 5 mJy (in a 3~\kms\ channel) on the existence of HC$_3$N (from this same transition) at the maser location.

A very significant difference between these two class~I maser hosts is the lack of 44.1-GHz emission observed towards NGC~4945. \citet{Ellingsen+17b} detected 44.1-GHz masers in two of the regions host to 36.2-GHz masers in NGC~4945. Despite the relative weakness of the 44.1-GHz emission in NGC~253, our 3$\sigma$ upper limit from NGC~4945 is a factor of two lower. Assuming our 5$\sigma$ upper limit (6~mJy peak) on the existence of 44.1-GHz emission, we place a lower limit on the 36.2:44.1~GHz integrated line intensity ratio of 15:1. Comparatively, the two regions in NGC~253 displaying emission from both transitions have ratios of 11:1 and 17:1 \citep[for regions `MM1' and `MM4' respectively;][]{Ellingsen+17b}, with lower limits on the regions without detections at 44.1-GHz ranging from 4:1 up to 18:1. Therefore, despite a potentially much lower peak flux density, 44.1-GHz emission may exist towards NGC~4945 at similar relative intensities as observed in NGC~253.

Given that the region of 36.2-GHz emission in NGC~4945 is significantly brighter than any of the regions in NGC~253 also hosting 44.1-GHz emission, we would expect any 44.1-GHz emission to be above this threshold assuming similar ratios to what is observed in NGC~253. 

Extended thermal methanol (48.4-GHz transition) is clearly visible in NGC~253, covering all sites where the 36.2-GHz maser emission is observed (Ellingsen et al., in preparation). A similar relationship is observed towards IC~342, with extended 96-GHz thermal methanol observed across all class~I maser locations \citep{Meier&05,Gorski+18}. Conversely, we see no thermal methanol from the 48.4-GHz transition towards NGC~4945. This, combined with differences in the 36.2- to 44.1-GHz flux density ratios and a  small line width (indicating a compact region), suggests that the optical depth of the masing line in NGC~4945 is likely much higher than in any region in NGC~253. 

\subsubsection{G\,1.6-0.025 }

G\,1.6-0.025 (referred to hereafter as G\,1.6) is a giant molecular cloud located at the eastern-most edge of the Milky Way's CMZ cloud complex \citep{Whiteoak+79, Bally+87, Bally+88}. This region displays maser emission from various class~I methanol transitions. However, unlike typical methanol maser sources, it lacks any significant high mass star-formation \citep{Haschick+93, Menten+09}. The nature of the maser emission from G\,1.6 is unlike that observed from typical Galactic class~I maser sites, and appears to be somewhat similar to those observed in the extragalactic class~I regions of NGC~4945 and NGC~253. Here we will make a comparison between G\,1.6 and these extragalactic sources, and discuss the possibility of them resulting from similar physical conditions.

Much like NGC~4945 (and to a lesser extent NGC~253), emission from the 44.1-GHz methanol transition is much weaker than the accompanying 36.2-GHz methanol emission in G\,1.6 \citep{Jones+13}. As previously mentioned, this is atypical for Galactic class~I maser sources, where these two transitions are often observed together, with the 44.1-GHz line being generally more luminous \citep{Voronkov+14}. This may indicate that G\,1.6 and these extragalactic class~I maser regions share a common pumping regime, which results in emission from the 36.2-GHz transition dominating. G\,1.6 is also host to maser emission from the 84.5-GHz methanol transition \citep{Salii+02}. This transition is related to the 36.2-GHz line, and is additionally observed tentatively as a megamaser in the starburst galaxy NGC~1068 \citep{Wang+14}. We suggest that it is likely that this 84.5-GHz methanol transition will be visible towards the extragalactic class~I regions in NGC~4945 and NGC~253. If future observations discover this to be correct, a comparison of flux densities between the 36.2- and 84.5-GHz lines in G\,1.6 and these extragalactic sources may reveal further similarities.

The broader environment of G\,1.6 also shares similarities with the extragalactic maser sources. G\,1.6 is a dense giant molecular cloud, easily visible in emission from the same dense gas tracers that identified similar clouds in NGC~253 \citep{Leroy+15}. Each independent region of 36.2-GHz maser emission from NGC~253 is associated with one of these giant molecular clouds \citep{Ellingsen+17b}. We also discuss in Section \ref{sec:methanol_maser}, that the south-eastern HCN and CS hotspot in NGC~4945 also indicates the presence of a giant molecular cloud at the location of the class~I maser emission \citep{Henkel+18}. In addition, the class~I masing regions of NGC~253 are bright in the 43.4-GHz SiO and 48.4-GHz thermal methanol transitions, which are some of the strongest 7-mm spectral lines observed towards G\,1.6 \citep{Jones+13}. However, these same transitions are not observed towards the masing region of NGC~4945, with emission from the 48.4-GHz thermal methanol not-detected, and weak 43.4-GHz SiO only observed close to the galactic nucleus.

One of the most interesting properties of G\,1.6 is the lack of star-formation from a region displaying such prominent 36.2-GHz class~I maser emission. However, this lack of star-formation does not appear to be mirrored in the environment of the class~I maser emission in NGC~4945, with enhanced star-formation (Knot B) observed nearby to the region of dense gas at the masers location \citep{Marconi+00, Henkel+18}. This star-formation is offset from the maser emission, however, it may be part of the same molecular cloud complex \citep{Henkel+18}. If an association with Knot B does exist, this would indicate that the class~I maser emission from NGC~4945 and G\,1.6 may not result from the same phenomenon and leaves open the possibility between a direct connection between the maser emission and star-formation.


\section{Conclusions}

Our high-resolution imaging follow up of the 36.2-GHz class~I methanol maser emission in NGC~4945 has confirmed its offset position from the galactic nucleus. Assuming the region is part of the disk, it would be located at a galactocentric radius of approximately 670~pc and is likely associated with the interface region between galactic bar and south-eastern spiral arm on the front side of the galaxy. This position corresponds to the position of a hotspot observed in dense gas tracers, likely indicating an association between the masing region and a giant molecular cloud. We detect methanol emission from neither the 44.1-GHz masing transion nor the 48.4-GHz thermal transition towards NGC~4945 and this indicates a high optical depth for the 36.2-GHz class~I masing region.

The 7-mm continuum source is offset by $0\farcs8\pm0\farcs4$ to the north-west of the dynamical centre of NGC~4945 ($\alpha_{2000} = 13^{\text{h}}~05^{\text{m}} ~ 27^{\text{s}}.467\pm0^{\text{s}}.032$ and $\delta_{2000} = -49^\circ ~ 28^\prime ~ 04\farcs8\pm0\farcs3$).

Emission from the HC$_3$N $J = 4 \rightarrow 3$, CS $J = 1 \rightarrow 0$ and SiO $J = 1 \rightarrow 0$ transitions were also detected towards NGC~4945. All of these thermal transitions were observed towards the central region of NGC~4945 (consistent with the 3-mm ALMA observations). Additionally, all of these species display strong absorption towards the continuum source, with a peak absorption component at 636~\kms. None of these molecular species were detected towards the offset location where the methanol masers are observed.

We identify many similarities between the class~I methanol masers of NGC~4945 and NGC~253. There is the possibility that the optical depth of the maser in NGC 4945 is higher, leading to particularly high intensity ratios between the 36.2- and 44.1-GHz methanol lines. A comparison with the giant molecular cloud G\,1.6-0.025 revealed similar line ratios between the 36.2- and 44.1-GHz methanol maser transitions, though unlike G\,1.6-0.025, the maser region in NGC~4945 may be associated with a region of enhanced star-formation. 

\section*{Acknowledgements}

We thank the referee for useful suggestions which helped to improve this paper. We thank the authors of \citet{Henkel+18} for allowing the use of their images cubes in the creation of our figures. The ATCA is part of the Australia Telescope which is funded by the Commonwealth of Australia for operation as a National Facility managed by CSIRO.  This research has made use of NASA's Astrophysics Data System Abstract Service. 
This research also utilised APLPY, an open-source plotting package for PYTHON hosted at http://aplpy.github.com. This research made use of Astropy, a community-developed core Python package for Astronomy \citep{astropy+13}.

\bibliography{references}

\newcommand{\noop}[1]{}
\begin{thebibliography}{}
\makeatletter
\relax
\def\mn@urlcharsother{\let\do\@makeother \do\$\do\&\do\#\do\^\do\_\do\%\do\~}
\def\mn@doi{\begingroup\mn@urlcharsother \@ifnextchar [ {\mn@doi@}
  {\mn@doi@[]}}
\def\mn@doi@[#1]#2{\def\@tempa{#1}\ifx\@tempa\@empty \href
  {http://dx.doi.org/#2} {doi:#2}\else \href {http://dx.doi.org/#2} {#1}\fi
  \endgroup}
\def\mn@eprint#1#2{\mn@eprint@#1:#2::\@nil}
\def\mn@eprint@arXiv#1{\href {http://arxiv.org/abs/#1} {{\tt arXiv:#1}}}
\def\mn@eprint@dblp#1{\href {http://dblp.uni-trier.de/rec/bibtex/#1.xml}
  {dblp:#1}}
\def\mn@eprint@#1:#2:#3:#4\@nil{\def\@tempa {#1}\def\@tempb {#2}\def\@tempc
  {#3}\ifx \@tempc \@empty \let \@tempc \@tempb \let \@tempb \@tempa \fi \ifx
  \@tempb \@empty \def\@tempb {arXiv}\fi \@ifundefined
  {mn@eprint@\@tempb}{\@tempb:\@tempc}{\expandafter \expandafter \csname
  mn@eprint@\@tempb\endcsname \expandafter{\@tempc}}}

\bibitem[\protect\citeauthoryear{{Astropy Collaboration} et~al.,}{{Astropy
  Collaboration} et~al.}{2013}]{astropy+13}
{Astropy Collaboration} et~al., 2013, \mn@doi [\aap]
  {10.1051/0004-6361/201322068}, \href
  {http://adsabs.harvard.edu/abs/2013A%26A...558A..33A} {558, A33}

\bibitem[\protect\citeauthoryear{{Bally}, {Stark}, {Wilson}  \&
  {Henkel}}{{Bally} et~al.}{1987}]{Bally+87}
{Bally} J.,  {Stark} A.~A.,  {Wilson} R.~W.,   {Henkel} C.,  1987, \mn@doi
  [\apjs] {10.1086/191217}, \href
  {http://adsabs.harvard.edu/abs/1987ApJS...65...13B} {65, 13}

\bibitem[\protect\citeauthoryear{{Bally}, {Stark}, {Wilson}  \&
  {Henkel}}{{Bally} et~al.}{1988}]{Bally+88}
{Bally} J.,  {Stark} A.~A.,  {Wilson} R.~W.,   {Henkel} C.,  1988, \mn@doi
  [\apj] {10.1086/165891}, \href
  {http://adsabs.harvard.edu/abs/1988ApJ...324..223B} {324, 223}

\bibitem[\protect\citeauthoryear{{Batrla}, {Matthews}, {Menten}  \&
  {Walmsley}}{{Batrla} et~al.}{1987}]{Batrla+87}
{Batrla} W.,  {Matthews} H.~E.,  {Menten} K.~M.,   {Walmsley} C.~M.,  1987,
  \mn@doi [\nat] {10.1038/326049a0}, \href
  {http://adsabs.harvard.edu/abs/1987Natur.326...49B} {326, 49}

\bibitem[\protect\citeauthoryear{{Bendo}, {Henkel}, {D'Cruze}, {Dickinson},
  {Fuller}  \& {Karim}}{{Bendo} et~al.}{2016}]{Bendo+16}
{Bendo} G.~J.,  {Henkel} C.,  {D'Cruze} M.~J.,  {Dickinson} C.,  {Fuller}
  G.~A.,   {Karim} A.,  2016, \mn@doi [\mnras] {10.1093/mnras/stw1659}, \href
  {http://adsabs.harvard.edu/abs/2016MNRAS.463..252B} {463, 252}

\bibitem[\protect\citeauthoryear{{Breen} et~al.,}{{Breen}
  et~al.}{2015}]{Breen+15}
{Breen} S.~L.,  et~al., 2015, \mn@doi [\mnras] {10.1093/mnras/stv847}, \href
  {http://adsabs.harvard.edu/abs/2015MNRAS.450.4109B} {450, 4109}

\bibitem[\protect\citeauthoryear{{Caswell} et~al.,}{{Caswell}
  et~al.}{2010}]{Caswell+10}
{Caswell} J.~L.,  et~al., 2010, \mn@doi [\mnras]
  {10.1111/j.1365-2966.2010.16339.x}, \href
  {http://adsabs.harvard.edu/abs/2010MNRAS.404.1029C} {404, 1029}

\bibitem[\protect\citeauthoryear{{Caswell} et~al.,}{{Caswell}
  et~al.}{2011}]{Caswell+11}
{Caswell} J.~L.,  et~al., 2011, \mn@doi [\mnras]
  {10.1111/j.1365-2966.2011.19383.x}, \href
  {http://adsabs.harvard.edu/abs/2011MNRAS.417.1964C} {417, 1964}

\bibitem[\protect\citeauthoryear{{Chen}, {Ellingsen}, {Shen}, {Titmarsh}  \&
  {Gan}}{{Chen} et~al.}{2011}]{Chen+11}
{Chen} X.,  {Ellingsen} S.~P.,  {Shen} Z.-Q.,  {Titmarsh} A.,   {Gan} C.-G.,
  2011, \mn@doi [\apjs] {10.1088/0067-0049/196/1/9}, \href
  {http://adsabs.harvard.edu/abs/2011ApJS..196....9C} {196, 9}

\bibitem[\protect\citeauthoryear{{Chen}, {Ellingsen}, {Baan}, {Qiao}, {Li},
  {An}  \& {Breen}}{{Chen} et~al.}{2015}]{Chen+15}
{Chen} X.,  {Ellingsen} S.~P.,  {Baan} W.~A.,  {Qiao} H.-H.,  {Li} J.,  {An}
  T.,   {Breen} S.~L.,  2015, \mn@doi [\apjl] {10.1088/2041-8205/800/1/L2},
  \href {http://adsabs.harvard.edu/abs/2015ApJ...800L...2C} {800, L2}

\bibitem[\protect\citeauthoryear{{Chen}, {Ellingsen}, {Shen}, {McCarthy},
  {Zhong}  \& {Deng}}{{Chen} et~al.}{2018}]{Chen+18}
{Chen} X.,  {Ellingsen} S.~P.,  {Shen} Z.-Q.,  {McCarthy} T.~P.,  {Zhong}
  W.-Y.,   {Deng} H.,  2018, \mn@doi [\apjl] {10.3847/2041-8213/aab894}, \href
  {http://adsabs.harvard.edu/abs/2018ApJ...856L..35C} {856, L35}

\bibitem[\protect\citeauthoryear{{Chou} et~al.,}{{Chou} et~al.}{2007}]{Chou+07}
{Chou} R.~C.~Y.,  et~al., 2007, \mn@doi [\apj] {10.1086/521351}, \href
  {http://adsabs.harvard.edu/abs/2007ApJ...670..116C} {670, 116}

\bibitem[\protect\citeauthoryear{{Cyganowski}, {Brogan}, {Hunter}  \&
  {Churchwell}}{{Cyganowski} et~al.}{2009}]{Cyganowski+09}
{Cyganowski} C.~J.,  {Brogan} C.~L.,  {Hunter} T.~R.,   {Churchwell} E.,  2009,
  \mn@doi [\apj] {10.1088/0004-637X/702/2/1615}, \href
  {http://adsabs.harvard.edu/abs/2009ApJ...702.1615C} {702, 1615}

\bibitem[\protect\citeauthoryear{{Cyganowski}, {Brogan}, {Hunter}, {Zhang},
  {Friesen}, {Indebetouw}  \& {Chandler}}{{Cyganowski}
  et~al.}{2012}]{Cyganowski+12}
{Cyganowski} C.~J.,  {Brogan} C.~L.,  {Hunter} T.~R.,  {Zhang} Q.,  {Friesen}
  R.~K.,  {Indebetouw} R.,   {Chandler} C.~J.,  2012, preprint, \href
  {http://adsabs.harvard.edu/abs/2012arXiv1210.3366C} {} (\mn@eprint {arXiv}
  {1210.3366})

\bibitem[\protect\citeauthoryear{{Dalcanton} et~al.,}{{Dalcanton}
  et~al.}{2009}]{Dalcanton+09}
{Dalcanton} J.~J.,  et~al., 2009, \mn@doi [\apjs] {10.1088/0067-0049/183/1/67},
  \href {http://adsabs.harvard.edu/abs/2009ApJS..183...67D} {183, 67}

\bibitem[\protect\citeauthoryear{{Ellingsen}, {Shabala}  \&
  {Kurtz}}{{Ellingsen} et~al.}{2005}]{Ellingsen+05}
{Ellingsen} S.~P.,  {Shabala} S.~S.,   {Kurtz} S.~E.,  2005, \mn@doi [\mnras]
  {10.1111/j.1365-2966.2005.08716.x}, \href
  {http://adsabs.harvard.edu/abs/2005MNRAS.357.1003E} {357, 1003}

\bibitem[\protect\citeauthoryear{{Ellingsen}, {Breen}, {Caswell}, {Quinn}  \&
  {Fuller}}{{Ellingsen} et~al.}{2010}]{Ellingsen+10}
{Ellingsen} S.~P.,  {Breen} S.~L.,  {Caswell} J.~L.,  {Quinn} L.~J.,   {Fuller}
  G.~A.,  2010, \mn@doi [\mnras] {10.1111/j.1365-2966.2010.16349.x}, \href
  {http://adsabs.harvard.edu/abs/2010MNRAS.404..779E} {404, 779}

\bibitem[\protect\citeauthoryear{{Ellingsen}, {Chen}, {Qiao}, {Baan}, {An},
  {Li}  \& {Breen}}{{Ellingsen} et~al.}{2014}]{Ellingsen+14}
{Ellingsen} S.~P.,  {Chen} X.,  {Qiao} H.-H.,  {Baan} W.,  {An} T.,  {Li} J.,
  {Breen} S.~L.,  2014, \mn@doi [\apjl] {10.1088/2041-8205/790/2/L28}, \href
  {http://adsabs.harvard.edu/abs/2014ApJ...790L..28E} {790, L28}

\bibitem[\protect\citeauthoryear{{Ellingsen}, {Chen}, {Breen}  \&
  {Qiao}}{{Ellingsen} et~al.}{2017a}]{Ellingsen+17b}
{Ellingsen} S.~P.,  {Chen} X.,  {Breen} S.~L.,   {Qiao} H.-H.,  2017a, \mn@doi
  [\mnras] {10.1093/mnras/stx2076}, \href
  {http://adsabs.harvard.edu/abs/2017MNRAS.472..604E} {472, 604}

\bibitem[\protect\citeauthoryear{{Ellingsen}, {Chen}, {Breen}  \&
  {Qiao}}{{Ellingsen} et~al.}{2017b}]{Ellingsen+17}
{Ellingsen} S.~P.,  {Chen} X.,  {Breen} S.~L.,   {Qiao} H.-h.,  2017b, \mn@doi
  [\apjl] {10.3847/2041-8213/aa71a6}, \href
  {http://adsabs.harvard.edu/abs/2017ApJ...841L..14E} {841, L14}

\bibitem[\protect\citeauthoryear{{Gan}, {Chen}, {Shen}, {Xu}  \& {Ju}}{{Gan}
  et~al.}{2013}]{Gan+13}
{Gan} C.-G.,  {Chen} X.,  {Shen} Z.-Q.,  {Xu} Y.,   {Ju} B.-G.,  2013, \mn@doi
  [\apj] {10.1088/0004-637X/763/1/2}, \href
  {http://adsabs.harvard.edu/abs/2013ApJ...763....2G} {763, 2}

\bibitem[\protect\citeauthoryear{{Gorski}, {Ott}, {Rand}, {Meier}, {Momjian}
  \& {Schinnerer}}{{Gorski} et~al.}{2018}]{Gorski+18}
{Gorski} M.,  {Ott} J.,  {Rand} R.,  {Meier} D.,  {Momjian} E.,   {Schinnerer}
  E.,  2018, preprint, \href
  {http://adsabs.harvard.edu/abs/2018arXiv180400578G} {} (\mn@eprint {arXiv}
  {1804.00578})

\bibitem[\protect\citeauthoryear{{Green} et~al.,}{{Green}
  et~al.}{2008}]{Green+08}
{Green} J.~A.,  et~al., 2008, \mn@doi [\mnras]
  {10.1111/j.1365-2966.2008.12888.x}, \href
  {http://adsabs.harvard.edu/abs/2008MNRAS.385..948G} {385, 948}

\bibitem[\protect\citeauthoryear{{Green} et~al.,}{{Green}
  et~al.}{2010}]{Green+10}
{Green} J.~A.,  et~al., 2010, \mn@doi [\mnras]
  {10.1111/j.1365-2966.2010.17376.x}, \href
  {http://adsabs.harvard.edu/abs/2010MNRAS.409..913G} {409, 913}

\bibitem[\protect\citeauthoryear{{Green} et~al.,}{{Green}
  et~al.}{2012}]{Green+12a}
{Green} J.~A.,  et~al., 2012, \mn@doi [\mnras]
  {10.1111/j.1365-2966.2011.20229.x}, \href
  {http://adsabs.harvard.edu/abs/2012MNRAS.420.3108G} {420, 3108}

\bibitem[\protect\citeauthoryear{{Green} et~al.,}{{Green}
  et~al.}{2017}]{Green+17}
{Green} J.~A.,  et~al., 2017, \mn@doi [\mnras] {10.1093/mnras/stx887}, \href
  {http://adsabs.harvard.edu/abs/2017MNRAS.469.1383G} {469, 1383}

\bibitem[\protect\citeauthoryear{{Greenhill}, {Moran}  \&
  {Herrnstein}}{{Greenhill} et~al.}{1997}]{Greenhill+97}
{Greenhill} L.~J.,  {Moran} J.~M.,   {Herrnstein} J.~R.,  1997, \mn@doi [\apjl]
  {10.1086/310643}, \href {http://adsabs.harvard.edu/abs/1997ApJ...481L..23G}
  {481, L23}

\bibitem[\protect\citeauthoryear{{Hagiwara}, {Horiuchi}, {Doi}, {Miyoshi}  \&
  {Edwards}}{{Hagiwara} et~al.}{2016}]{Hagiwara+16}
{Hagiwara} Y.,  {Horiuchi} S.,  {Doi} A.,  {Miyoshi} M.,   {Edwards} P.~G.,
  2016, \mn@doi [\apj] {10.3847/0004-637X/827/1/69}, \href
  {http://adsabs.harvard.edu/abs/2016ApJ...827...69H} {827, 69}

\bibitem[\protect\citeauthoryear{{Haschick} \& {Baan}}{{Haschick} \&
  {Baan}}{1993}]{Haschick+93}
{Haschick} A.~D.,  {Baan} W.~A.,  1993, \mn@doi [\apj] {10.1086/172782}, \href
  {http://adsabs.harvard.edu/abs/1993ApJ...410..663H} {410, 663}

\bibitem[\protect\citeauthoryear{{Henkel} et~al.,}{{Henkel}
  et~al.}{2018}]{Henkel+18}
{Henkel} C.,  et~al., 2018, preprint, \href
  {http://adsabs.harvard.edu/abs/2018arXiv180209852H} {} (\mn@eprint {arXiv}
  {1802.09852})

\bibitem[\protect\citeauthoryear{{Humphreys}, {Vlemmings}, {Impellizzeri},
  {Galametz}, {Olberg}, {Conway}, {Belitsky}  \& {De Breuck}}{{Humphreys}
  et~al.}{2016}]{Humphreys+16}
{Humphreys} E.~M.~L.,  {Vlemmings} W.~H.~T.,  {Impellizzeri} C.~M.~V.,
  {Galametz} M.,  {Olberg} M.,  {Conway} J.~E.,  {Belitsky} V.,   {De Breuck}
  C.,  2016, \mn@doi [\aap] {10.1051/0004-6361/201629168}, \href
  {http://adsabs.harvard.edu/abs/2016A%26A...592L..13H} {592, L13}

\bibitem[\protect\citeauthoryear{{Hunt}, {Whiteoak}, {Cragg}, {White}  \&
  {Jones}}{{Hunt} et~al.}{1999}]{Hunt+99}
{Hunt} M.~R.,  {Whiteoak} J.~B.,  {Cragg} D.~M.,  {White} G.~L.,   {Jones}
  P.~A.,  1999, \mn@doi [\mnras] {10.1046/j.1365-8711.1999.01872.x}, \href
  {http://adsabs.harvard.edu/abs/1999MNRAS.302....1H} {302, 1}

\bibitem[\protect\citeauthoryear{{Ishihara}, {Nakai}, {Iyomoto}, {Makishima},
  {Diamond}  \& {Hall}}{{Ishihara} et~al.}{2001}]{Ishihara+01}
{Ishihara} Y.,  {Nakai} N.,  {Iyomoto} N.,  {Makishima} K.,  {Diamond} P.,
  {Hall} P.,  2001, \mn@doi [\pasj] {10.1093/pasj/53.2.215}, \href
  {http://adsabs.harvard.edu/abs/2001PASJ...53..215I} {53, 215}

\bibitem[\protect\citeauthoryear{{Jones}, {Burton}, {Cunningham}, {Tothill}  \&
  {Walsh}}{{Jones} et~al.}{2013}]{Jones+13}
{Jones} P.~A.,  {Burton} M.~G.,  {Cunningham} M.~R.,  {Tothill} N.~F.~H.,
  {Walsh} A.~J.,  2013, \mn@doi [\mnras] {10.1093/mnras/stt717}, \href
  {http://adsabs.harvard.edu/abs/2013MNRAS.433..221J} {433, 221}

\bibitem[\protect\citeauthoryear{{Jordan}, {Walsh}, {Breen}, {Ellingsen},
  {Voronkov}  \& {Hyland}}{{Jordan} et~al.}{2017}]{Jordan+17}
{Jordan} C.~H.,  {Walsh} A.~J.,  {Breen} S.~L.,  {Ellingsen} S.~P.,  {Voronkov}
  M.~A.,   {Hyland} L.~J.,  2017, \mn@doi [\mnras] {10.1093/mnras/stx1776},
  \href {http://adsabs.harvard.edu/abs/2017MNRAS.471.3915J} {471, 3915}

\bibitem[\protect\citeauthoryear{{Kurtz}, {Hofner}  \& {{\'A}lvarez}}{{Kurtz}
  et~al.}{2004}]{Kurtz+04}
{Kurtz} S.,  {Hofner} P.,   {{\'A}lvarez} C.~V.,  2004, \mn@doi [\apjs]
  {10.1086/423956}, \href {http://adsabs.harvard.edu/abs/2004ApJS..155..149K}
  {155, 149}

\bibitem[\protect\citeauthoryear{{Leroy} et~al.,}{{Leroy}
  et~al.}{2015}]{Leroy+15}
{Leroy} A.~K.,  et~al., 2015, \mn@doi [\apj] {10.1088/0004-637X/801/1/25},
  \href {http://adsabs.harvard.edu/abs/2015ApJ...801...25L} {801, 25}

\bibitem[\protect\citeauthoryear{{Licquia} \& {Newman}}{{Licquia} \&
  {Newman}}{2015}]{Licquia+15}
{Licquia} T.~C.,  {Newman} J.~A.,  2015, \mn@doi [\apj]
  {10.1088/0004-637X/806/1/96}, \href
  {http://adsabs.harvard.edu/abs/2015ApJ...806...96L} {806, 96}

\bibitem[\protect\citeauthoryear{{Marconi}, {Oliva}, {van der Werf},
  {Maiolino}, {Schreier}, {Macchetto}  \& {Moorwood}}{{Marconi}
  et~al.}{2000}]{Marconi+00}
{Marconi} A.,  {Oliva} E.,  {van der Werf} P.~P.,  {Maiolino} R.,  {Schreier}
  E.~J.,  {Macchetto} F.,   {Moorwood} A.~F.~M.,  2000, \aap, \href
  {http://adsabs.harvard.edu/abs/2000A\%26A...357...24M} {357, 24}

\bibitem[\protect\citeauthoryear{{McCarthy}, {Ellingsen}, {Chen}, {Breen},
  {Voronkov}  \& {Qiao}}{{McCarthy} et~al.}{2017}]{McCarthy+17}
{McCarthy} T.~P.,  {Ellingsen} S.~P.,  {Chen} X.,  {Breen} S.~L.,  {Voronkov}
  M.~A.,   {Qiao} H.-h.,  2017, \mn@doi [\apj] {10.3847/1538-4357/aa872c},
  \href {http://adsabs.harvard.edu/abs/2017ApJ...846..156M} {846, 156}

\bibitem[\protect\citeauthoryear{{McGee}, {Newton}  \& {Balister}}{{McGee}
  et~al.}{1977}]{McGee+77}
{McGee} R.~X.,  {Newton} L.~M.,   {Balister} M.,  1977, \mn@doi [\mnras]
  {10.1093/mnras/180.4.585}, \href
  {http://adsabs.harvard.edu/abs/1977MNRAS.180..585M} {180, 585}

\bibitem[\protect\citeauthoryear{{Meier} \& {Turner}}{{Meier} \&
  {Turner}}{2005}]{Meier&05}
{Meier} D.~S.,  {Turner} J.~L.,  2005, \mn@doi [\apj] {10.1086/426499}, \href
  {http://adsabs.harvard.edu/abs/2005ApJ...618..259M} {618, 259}

\bibitem[\protect\citeauthoryear{{Meier} et~al.,}{{Meier}
  et~al.}{2015}]{Meier+15}
{Meier} D.~S.,  et~al., 2015, \mn@doi [\apj] {10.1088/0004-637X/801/1/63},
  \href {http://adsabs.harvard.edu/abs/2015ApJ...801...63M} {801, 63}

\bibitem[\protect\citeauthoryear{{Menten}}{{Menten}}{1991}]{Menten91a}
{Menten} K.,  1991, in {Haschick} A.~D.,  {Ho} P.~T.~P.,  eds,  {Astronomical
  Society of the Pacific Conference Series} Vol. 16, {Atoms, Ions and
  Molecules: New Results in Spectral Line Astrophysics}. p.~119

\bibitem[\protect\citeauthoryear{{Menten}, {Wilson}, {Leurini}  \&
  {Schilke}}{{Menten} et~al.}{2009}]{Menten+09}
{Menten} K.~M.,  {Wilson} R.~W.,  {Leurini} S.,   {Schilke} P.,  2009, \mn@doi
  [\apj] {10.1088/0004-637X/692/1/47}, \href
  {http://adsabs.harvard.edu/abs/2009ApJ...692...47M} {692, 47}

\bibitem[\protect\citeauthoryear{{Moorwood}, {van der Werf}, {Kotilainen},
  {Marconi}  \& {Oliva}}{{Moorwood} et~al.}{1996}]{Moorwood+96}
{Moorwood} A.~F.~M.,  {van der Werf} P.~P.,  {Kotilainen} J.~K.,  {Marconi} A.,
    {Oliva} E.,  1996, \aap, \href
  {http://adsabs.harvard.edu/abs/1996A%26A...308L...1M} {308, L1}

\bibitem[\protect\citeauthoryear{{Ott}, {Whiteoak}, {Henkel}  \&
  {Wielebinski}}{{Ott} et~al.}{2001}]{Ott+01}
{Ott} M.,  {Whiteoak} J.~B.,  {Henkel} C.,   {Wielebinski} R.,  2001, \mn@doi
  [\aap] {10.1051/0004-6361:20010505}, \href
  {http://adsabs.harvard.edu/abs/2001A\%26A...372..463O} {372, 463}

\bibitem[\protect\citeauthoryear{{P{\'e}rez-Beaupuits}, {Spoon}, {Spaans}  \&
  {Smith}}{{P{\'e}rez-Beaupuits} et~al.}{2011}]{Perez-Beauputis+11}
{P{\'e}rez-Beaupuits} J.~P.,  {Spoon} H.~W.~W.,  {Spaans} M.,   {Smith} J.~D.,
  2011, \mn@doi [\aap] {10.1051/0004-6361/201117153}, \href
  {http://adsabs.harvard.edu/abs/2011A\%26A...533A..56P} {533, A56}

\bibitem[\protect\citeauthoryear{{Pesce}, {Braatz}  \& {Impellizzeri}}{{Pesce}
  et~al.}{2016}]{Pesce+16}
{Pesce} D.~W.,  {Braatz} J.~A.,   {Impellizzeri} C.~M.~V.,  2016, \mn@doi
  [\apj] {10.3847/0004-637X/827/1/68}, \href
  {http://adsabs.harvard.edu/abs/2016ApJ...827...68P} {827, 68}

\bibitem[\protect\citeauthoryear{{Reid} et~al.,}{{Reid} et~al.}{2009}]{Reid+09}
{Reid} M.~J.,  et~al., 2009, \mn@doi [\apj] {10.1088/0004-637X/700/1/137},
  \href {http://adsabs.harvard.edu/abs/2009ApJ...700..137R} {700, 137}

\bibitem[\protect\citeauthoryear{{Sakamoto}, {Mao}, {Matsushita}, {Peck},
  {Sawada}  \& {Wiedner}}{{Sakamoto} et~al.}{2011}]{Sakamoto+11}
{Sakamoto} K.,  {Mao} R.-Q.,  {Matsushita} S.,  {Peck} A.~B.,  {Sawada} T.,
  {Wiedner} M.~C.,  2011, \mn@doi [\apj] {10.1088/0004-637X/735/1/19}, \href
  {http://adsabs.harvard.edu/abs/2011ApJ...735...19S} {735, 19}

\bibitem[\protect\citeauthoryear{{Salii}, {Sobolev}  \& {Kalinina}}{{Salii}
  et~al.}{2002}]{Salii+02}
{Salii} S.~V.,  {Sobolev} A.~M.,   {Kalinina} N.~D.,  2002, \mn@doi [Astronomy
  Reports] {10.1134/1.1529254}, \href
  {http://adsabs.harvard.edu/abs/2002ARep...46..955S} {46, 955}

\bibitem[\protect\citeauthoryear{{Sjouwerman}, {Murray}, {Pihlstr{\"o}m},
  {Fish}  \& {Araya}}{{Sjouwerman} et~al.}{2010}]{Sjouwerman+10}
{Sjouwerman} L.~O.,  {Murray} C.~E.,  {Pihlstr{\"o}m} Y.~M.,  {Fish} V.~L.,
  {Araya} E.~D.,  2010, \mn@doi [\apjl] {10.1088/2041-8205/724/2/L158}, \href
  {http://adsabs.harvard.edu/abs/2010ApJ...724L.158S} {724, L158}

\bibitem[\protect\citeauthoryear{{Slysh}, {Kalenskii}, {Valtts}  \&
  {Otrupcek}}{{Slysh} et~al.}{1994}]{Slysh+94}
{Slysh} V.~I.,  {Kalenskii} S.~V.,  {Valtts} I.~E.,   {Otrupcek} R.,  1994,
  \mn@doi [\mnras] {10.1093/mnras/268.2.464}, \href
  {http://adsabs.harvard.edu/abs/1994MNRAS.268..464S} {268, 464}

\bibitem[\protect\citeauthoryear{{Spoon}, {Koornneef}, {Moorwood}, {Lutz}  \&
  {Tielens}}{{Spoon} et~al.}{2000}]{Spoon+00}
{Spoon} H.~W.~W.,  {Koornneef} J.,  {Moorwood} A.~F.~M.,  {Lutz} D.,
  {Tielens} A.~G.~G.~M.,  2000, \aap, \href
  {http://adsabs.harvard.edu/abs/2000A\%26A...357..898S} {357, 898}

\bibitem[\protect\citeauthoryear{{Spoon}, {Moorwood}, {Pontoppidan}, {Cami},
  {Kregel}, {Lutz}  \& {Tielens}}{{Spoon} et~al.}{2003}]{Spoon+03}
{Spoon} H.~W.~W.,  {Moorwood} A.~F.~M.,  {Pontoppidan} K.~M.,  {Cami} J.,
  {Kregel} M.,  {Lutz} D.,   {Tielens} A.~G.~G.~M.,  2003, \mn@doi [\aap]
  {10.1051/0004-6361:20030290}, \href
  {http://adsabs.harvard.edu/abs/2003A\%26A...402..499S} {402, 499}

\bibitem[\protect\citeauthoryear{{Strickland}, {Heckman}, {Colbert}, {Hoopes}
  \& {Weaver}}{{Strickland} et~al.}{2004}]{Strickland+04}
{Strickland} D.~K.,  {Heckman} T.~M.,  {Colbert} E.~J.~M.,  {Hoopes} C.~G.,
  {Weaver} K.~A.,  2004, \mn@doi [\apj] {10.1086/383136}, \href
  {http://adsabs.harvard.edu/abs/2004ApJ...606..829S} {606, 829}

\bibitem[\protect\citeauthoryear{{Tully} et~al.,}{{Tully}
  et~al.}{2013}]{Tully+13}
{Tully} R.~B.,  et~al., 2013, \mn@doi [\aj] {10.1088/0004-6256/146/4/86}, \href
  {http://adsabs.harvard.edu/abs/2013AJ....146...86T} {146, 86}

\bibitem[\protect\citeauthoryear{{Val'tts}, {Ellingsen}, {Slysh}, {Kalenskii},
  {Otrupcek}  \& {Larionov}}{{Val'tts} et~al.}{2000}]{Valtts+00}
{Val'tts} I.~E.,  {Ellingsen} S.~P.,  {Slysh} V.~I.,  {Kalenskii} S.~V.,
  {Otrupcek} R.,   {Larionov} G.~M.,  2000, \mn@doi [\mnras]
  {10.1046/j.1365-8711.2000.03518.x}, \href
  {http://adsabs.harvard.edu/abs/2000MNRAS.317..315V} {317, 315}

\bibitem[\protect\citeauthoryear{{Voronkov}, {Caswell}, {Ellingsen}  \&
  {Sobolev}}{{Voronkov} et~al.}{2010}]{Voronkov+10a}
{Voronkov} M.~A.,  {Caswell} J.~L.,  {Ellingsen} S.~P.,   {Sobolev} A.~M.,
  2010, \mn@doi [\mnras] {10.1111/j.1365-2966.2010.16624.x}, \href
  {http://adsabs.harvard.edu/abs/2010MNRAS.405.2471V} {405, 2471}

\bibitem[\protect\citeauthoryear{{Voronkov}, {Caswell}, {Ellingsen}, {Green}
  \& {Breen}}{{Voronkov} et~al.}{2014}]{Voronkov+14}
{Voronkov} M.~A.,  {Caswell} J.~L.,  {Ellingsen} S.~P.,  {Green} J.~A.,
  {Breen} S.~L.,  2014, \mn@doi [\mnras] {10.1093/mnras/stu116}, \href
  {http://adsabs.harvard.edu/abs/2014MNRAS.439.2584V} {439, 2584}

\bibitem[\protect\citeauthoryear{{Wang}, {Zhang}, {Gao}, {Zhang}, {Li}, {Fang}
  \& {Shi}}{{Wang} et~al.}{2014}]{Wang+14}
{Wang} J.,  {Zhang} J.,  {Gao} Y.,  {Zhang} Z.-Y.,  {Li} D.,  {Fang} M.,
  {Shi} Y.,  2014, \mn@doi [Nature Communications] {10.1038/ncomms6449}, \href
  {http://adsabs.harvard.edu/abs/2014NatCo...5E5449W} {5, 5449}

\bibitem[\protect\citeauthoryear{{Whiteoak} \& {Gardner}}{{Whiteoak} \&
  {Gardner}}{1979}]{Whiteoak+79}
{Whiteoak} J.~B.,  {Gardner} F.~F.,  1979, \mn@doi [\mnras]
  {10.1093/mnras/188.3.445}, \href
  {http://adsabs.harvard.edu/abs/1979MNRAS.188..445W} {188, 445}

\bibitem[\protect\citeauthoryear{{Wilson} et~al.,}{{Wilson}
  et~al.}{2011}]{Wilson+11}
{Wilson} W.~E.,  et~al., 2011, \mn@doi [\mnras]
  {10.1111/j.1365-2966.2011.19054.x}, \href
  {http://adsabs.harvard.edu/abs/2011MNRAS.416..832W} {416, 832}

\makeatother
\end{thebibliography}

\onecolumn

\appendix

\end{document}